\documentclass{article}
\usepackage[a4paper, total={6in, 8in}]{geometry}
\usepackage{amsmath}
\usepackage{amsthm}
\usepackage{xcolor}
\usepackage{amssymb}
\usepackage{graphics}
\usepackage{mathtools}
\usepackage{empheq}
\usepackage{caption}
\usepackage{subcaption}
\DeclareMathOperator{\tr}{tr}
\newtheorem*{remark}{Remark}

\begin{document}

\begin{center}
    {\Large\bfseries Chemomechanical regulation of growing tissues from a thermodynamically-consistent framework and its application to tumor spheroid growth}\\[1em]
    {\large Nonthakorn Olaranont$^{1*}$, Chaozhen Wei$^{2*}$, John Lowengrub$^{1,3,4,5,6}$ and
Min Wu$^{7,8}$}\\
    {\small $^{1}$Department of Mathematics, University of California Irvine, Irvine, CA 92697 USA\\
    $^{2}$School of Mathematical Sciences, University of Electronic Science and Technology of China, Chengdu, Sichuan 611731, China\\
$^{3}$Department of Biomedical Engineering, University of California Irvine, Irvine, CA 92697 USA\\
$^{4}$Center for Multiscale Cell Fate Studies, University of California Irvine, Irvine, CA 92697 USA\\
$^{5}$Center for Complex Biological Systems, University of California Irvine, Irvine, CA 92697 USA\\
$^{6}$Chao Family Comprehensive Cancer Center, University of California, Irvine, CA 92697 USA\\
$^{7}$Department of Mathematical Sciences, Worcester Polytechnic Institute, Worcester, MA 01609 USA\\
$^{8}$Center for Computational Biology, Flatiron Institute, New York, NY 10010 USA\\
$^{*}$NO and CW have contributed equally to this work.}\\[2em]
\end{center}

\rule{\textwidth}{0.5pt}
\section*{\centering Abstract}
It is widely recognized that reciprocal interactions between cells and their microenvironment, via mechanical forces and biochemical signaling pathways, regulate cell behaviors during normal development, homeostasis and disease progression such as cancer. However, it is still not well understood how complex patterns of tissue growth emerge. Here, we propose a framework for the chemomechanical regulation of growth based on thermodynamics of continua and growth-elasticity to predict growth patterns. Combining the elastic and chemical energies, we use an energy variational approach to derive a novel formulation that incorporates an energy-dissipating stress relaxation and biochemomechanical regulation of the volumetric growth rate. We validate the model using experimental data from growth of tumor spheroids in confined environments. We also investigate the influence of model parameters, including tissue rearrangement rate, tissue compressibility, strength of mechanical feedback and external mechanical stimuli, on the growth patterns of tumor spheroids.

\vspace{1.5em} 
\noindent \textbf{Subjects:} Biomechanics, Mathematical modeling, Computational biology.

\noindent \textbf{Keywords:} growth, viscoelasticity, mechanical feedback, fluidity, nonlinear mechanics, tumor.

\noindent \textbf{Corresponding Author:} 
John Lowengrub, lowengrb@math.uci.edu;
Min Wu, englier@gmail.com


\newpage

\section{Introduction}
Developmental processes \cite{neufeld1998coordination,lecuit2003developmental,ingber2006mechanical}, tissue regeneration \cite{loewenstein1967intercellular,sun2014control}, cancer progression \cite{folkman2002role,altorki2019lung}, and organoid engineering \cite{lancaster2014organogenesis,hofer2021engineering} all involve growth of tissues and the complex interplay between biochemical signaling and mechanical forces \cite{thomson1917growth}. Understanding the dynamics of these multiscale processes requires elucidating the interactions among proliferating and non-proliferating cells, interstitial fluid and the extracellular matrix in a microenvironmental milieu \cite{hallatschek2023proliferating}. At the tissue or continuum level, growth involves mass being incorporated into a pre-existing living system at specific local rates
\cite{pan2016differential,tozluoglu2019planar} and in certain directions
\cite{thery2005extracellular,segalen2009cell,van2020oriented}. 

Taking tumor tissue for example, Greenspan assumed that tissue masses are nearly incompressible, and that mass is incorporated via cell proliferation under a rate proportional to the local “nutrient” level while counterbalancing a rate of mass removal, attributed to cell death \cite{greenspan1976growth}. This description connects the volumetric growth of mass with underlying cell birth and death processes phenomenologically in response to a chemical field. Although it does not detail cell-level growth regulation, it effectively describes the growth process of the living tissues at large-scales and can generate intricate tumor morphologies that resemble experimental observations \cite{lowengrub2009nonlinear}. 

Growth generates forces \cite{hallatschek2023proliferating} that arise from cells pushing and pulling on one another and the extracellular matrix, from active contractilities and from boundary conditions. However, the spatial transmission and persistence of these forces are determined by the material properties of the tissue constituents \cite{lecuit2011force}. To maintain tissue integrity, the push-and-pull between the cells and extracellular matrix is transmitted through the tissue as elastic forces, which can dissipate due to cell and subcellular rearrangements \cite{doubrovinski2017measurement},  as well as extracellular matrix remodeling \cite{chaudhuri2020effects}. When the dissipation is fast enough, tissue mechanical properties have been approximated by viscoelastic or viscous fluids \cite{Ranft-2010-pnas,Streichan-2018-elife}. When dissipation is slow, the deformation of tissues is purely elastic thus residual stresses can be retained even after the tissue ceases to grow \cite{nia2016solid}. 

Along these lines, Rodriguez et al. \cite{Rodriguez1994} developed a growth-elasticity theory using a configurative approach (aka morphoelasticity \cite{Goriely-2005-prl,goriely2007role,goriely2017mathematics,walker2023}), which has been widely used to describe various growth and morphogenesis processes (see the reviews \cite{taber1995biomechanics,ambrosi2019growth}). More recently, viscoelastic models  have been developed to study stress generation, transmission and elastic relaxation on tumor growth
\cite{ambrosi2009cell,yan2021stress,Wei2023elasticmodel,garcke2022,garcke2024}. For instance, in \cite{Wei2023elasticmodel,yan2021stress}, we merged growth-elasticity theory with an adaptive reference map  \cite{kamrin2012reference,rycroft2020reference} in which a phenomenological parameter, the rearrangement rate, is introduced to describe how fast the memory of the initial tissue configuration is lost. In this theory, tissues can behave like viscoelastic or viscous fluids as in \cite{Ranft-2010-pnas,Streichan-2018-elife} with positive rearrangement rate and like elastic solids when the rearrangement rate is small \cite{Rodriguez1994}. Recently, an alternative that incorporates fundamental thermodynamics concepts and general viscoelastic constitutive equations was recently developed, analyzed and simulated numerically in the context of phase field models of tumor growth  \cite{garcke2022,garcke2024}.

Mechanical stresses also regulate growth \cite{shraiman2005mechanical}, and this regulation is critical to both developmental processes \cite{pan2016differential,parada2022mechanical} and homeostasis \cite{humphrey2014mechanotransduction,eichinger2021mechanical}. While the standard between “normal” vs “abnormal” mechanical regulation of growth has not been well established, evidence shows that the tumor spheroids of various cell types grow with different rates under different levels of external compression or spatial confinement \cite{helmlinger1997solid,Montel2021}. Mathematical models have now been developed that incorporate mechanical feedback on growth processes \cite{araujo2004linear,budday2015size,garcia2017contraction,yan2021stress,Wei2023elasticmodel,garcke2024}. For example, in \cite{yan2021stress,Wei2023elasticmodel}, we regulated nutrient-driven growth using mechanical feedback driven by local compressive stresses through Michaelis–Menten kinetics. Although we found the model-predicted tumor growth fits the experimental data well \cite{yan2021stress}, we note that the mechanical feedback mechanisms considered by us and others are still open to debate and refinement.

In this work, we develop a new model framework for the chemomechanical regulation of growth. Different from our previous approach, here we obtain the forms of chemomechanical regulation and dissipative rearrangement by appealing to thermodynamic consistency \cite{gurtin2010mechanics} in the theory of growth-elasticity. In particular, we assume that the trajectory of the growth tensor field, or equivalently the elastic deformation field, dissipates the internal energy of a controlled system comprised of a growing continuum immersed a biochemical, growth-promoting field (e.g., nutrient). By ignoring the kinetic energy and temperature effects, the chemical energy from the nutrient field and the elastic energy from the tissue make up the total internal energy. 
By assuming mass-volumetric tissue growth is as an energy-dissipative process that converts the chemical energy into elastic energy, the mathematical form of the chemomechanical feedback function emerges naturally. By assuming mass-conserving tissue rearrangement to be energy dissipative, we obtain a stress dissipation process similar to the Maxwell-type fluid model. These energy-dissipative processes, the emergence of a natural form of chemomechanical feedback, and local mass conservation distinguish our approach from the thermodynamic, viscoelastic model developed in \cite{garcke2022,garcke2024}.

We apply the new model to study tumor spheroid growth, reproducing the experimental data on spheroid-radius dynamics in various scenarios, such as growth in a gel-constrained environment
 \cite{helmlinger1997solid} and growth with external compression applied on the spheroid surface \cite{Montel2021}. We further investigate the model parameters and show that the new chemomechanical regulation leads to several distinct predictions than seen in previous works. Most importantly, it predicts that the tumor spheroid continues to grow without external mechanical loads, in contrast to previous models where a steady-state size will always be reached. In addition, it predicts smaller local tissue rearrangement rates result in more uniform growth-rate patterns.

The paper is organized as follows. In Sec. 2, we lay out the fundamentals on the growth-elasticity theory in an Eulerian reference frame. In Sec. 3, we derive a new form of chemomechanical regulation within the growth-elasticity framework and compare the new model with previous growth models with mechanical feedback. In Sec. 4, we develop a numerical method to solve the new system in a reduced radially-symmetric geometry. In Sec. 5, we present the results of numerical simulations of tumor spheroid growth in constrained environments. In Sec. 6, we discuss the results and compare with previous theories and experiments.

\section{Growth-Elasticity in an Eulerian Frame}

In this section, we describe the decomposition of the deformation gradient, its dynamic behavior, the principles governing local growth and mass conservation, and the conditions governing neo-Hookean elasticity. We present the model equations in the Eulerian frame in preparation for coupling with the reaction-diffusion processes for the biochemical (nutrient) field, which is naturally described in the Eulerian coordinate system. 

\subsection{Decomposition of the Deformation Gradient}
\begin{figure}[!h]
	\centering
	\includegraphics[width=0.5\textwidth]{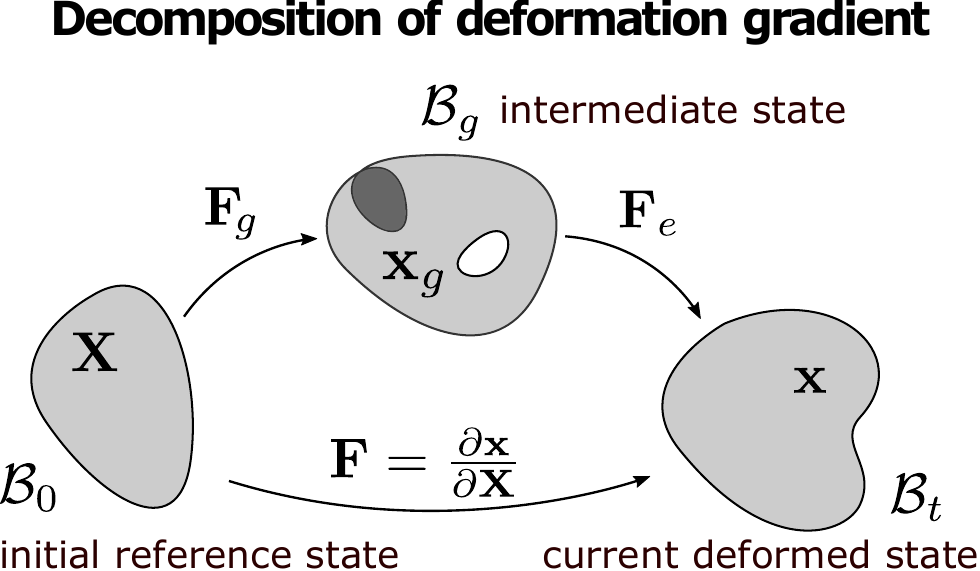}
	\caption{Decomposition of the geometric deformation gradient $\mathbf{F}=\mathbf{F}_e \mathbf{F}_g$. Through the two-stage transform $\mathcal{B}_0\rightarrow\mathcal{B}_g\rightarrow\mathcal{B}_t$, the tissue transforms from the initial reference state $\mathbf{X} \in \mathcal{B}_0$ to an intermediate state $\mathbf{x}_g \in \mathcal{B}_g$ by the growth stretch tensor $\mathbf{F}_g$, and then transforms to the current deformed state $\mathbf{x} \in \mathcal{B}_t$ by the elastic deformation tensor $\mathbf{F}_e$.  }\label{fig:decomposition_map}
\end{figure}

We decompose the total deformation of the growing tissue into two stages (as shown in Fig.~\ref{fig:decomposition_map}): at the first stage of growth, a material point at the initial state $\mathbf{X}\in\mathcal{B}_0 (\subset R^d)$ transforms to an intermediate state $\mathbf{x}_g \in \mathcal{B}_g$; then at the second state of elastic deformation, the material point transforms from $\mathbf{x}_g$ to a new position at the current deformed state $\mathbf{x} \in \mathcal{B}_t$. By the theory of finite elasticity and growth \cite{Rodriguez1994}, this decomposition can be described by
\begin{equation}
\label{decomposition}
	\mathbf{F}= \mathbf{F}_e \mathbf{F}_g
\end{equation}
where the geometric deformation gradient $\mathbf{F}=\partial \mathbf{x}/\partial \mathbf{X}$ is decomposed multiplicatively into the accumulated growth stretch tensor $\mathbf{F}_g$ and the elastic deformation tensor $\mathbf{F}_e$. Here we assume that the active growth only induces (possibly incompatible) changes in size and shape of the local continuum of existing materials without adding new material points or introducing stresses; while the elastic deformation consists of the necessary changes for maintaining continuum compatibility and the response to external loading or geometric constraints from the boundary. 

\subsection{Dynamics of the Deformation Gradient}
Since the initial reference coordinates of a material point do not change in time, we have $d\mathbf{X}/dt=0$, where $d\Box/dt = \partial \Box(\mathbf{X},t)/\partial t$ is the material time derivative. Taking the material time derivative of the deformation gradient $\mathbf{F}$, we obtain
\begin{equation}\label{eq:F_evol}
\frac{d \mathbf{F}}{dt} = \frac{d}{dt}\Big(\frac{\partial \mathbf{x}}{\partial \mathbf{X}}\Big)= \frac{\partial }{\partial \mathbf{X}}\Big(\frac{d \mathbf{x}}{dt}\Big) = \frac{\partial \mathbf{v}}{\partial \mathbf{X}}=\frac{\partial \mathbf{v}}{\partial \mathbf{x}}\frac{\partial \mathbf{x}}{\partial \mathbf{X}}=\nabla \mathbf{v}\mathbf{F},
\end{equation}
where $\mathbf{v}=\frac{\partial}{\partial t}\mathbf{x}(\mathbf{X},t)$ is the material velocity. Denoting the local volumetric variation $J=\det(\mathbf{F})$, its evolution is given by 
\begin{equation}\label{eq:J_evol}
\frac{dJ}{dt} =\frac{\partial J}{\partial{t}}+ {\mathbf v} \cdot {\nabla} {J}= J\nabla\cdot\mathbf{v},
\end{equation}
where $\nabla\cdot\mathbf{v}$ represents the local volumetric expansion rate. In the Eulerian frame, we emphasize that as long as $\mathbf{v}$ is solved, $\mathbf{F}$ and $J$ can be further computed using the above two equations. This is in contrast to Lagrangian models where $\mathbf{x}(\mathbf{X},t)$ and the associated $\mathbf{F}$ need to be solved. 
  
\subsection{Active Growth and Mass Conservation}
Letting $\tilde{\mathbf{\boldsymbol\Gamma}}$ be the growth rate tensor defined in the intermediate state $\mathcal{B}_g$, the dynamics of growth stretch tensor is 
\begin{equation}\label{eq:Fg_evol}
\frac{d \mathbf{F}_g}{dt} = \tilde{\mathbf{\boldsymbol\Gamma}}\mathbf{F}_g.
\end{equation}
The growth rate tensor $\tilde{\boldsymbol{\Gamma}}$ in general can be decomposed into the isotropic part and deviatoric part 
\begin{equation}\label{eq:Gamma_decomp}
\tilde{\boldsymbol{\Gamma}} = \frac{\gamma}{d}\mathbf{I}+\tilde{\boldsymbol{\Gamma}}_{D},
\end{equation}
where $\gamma=\tr(\tilde{\boldsymbol{\Gamma}})$ is the volumetric growth rate and ${\tilde{\boldsymbol{\Gamma}}}_D := \tilde{\boldsymbol{\Gamma}} - \frac{\tr{\tilde{\boldsymbol{\Gamma}}}}{d}\mathbf{I}$ is the deviatoric growth rate tensor, which is traceless, i.e., $\tr(\tilde{\boldsymbol{\Gamma}}_{D})=0$. Note that the dynamics of the growth stretch tensor is frame invariant (see SM. ~\ref{FrameInvariance} for details). In general, the volumetric growth rate $\gamma$ can be spatially heterogeneous due to chemical and mechanical regulation, while $\tilde{\boldsymbol{\Gamma}}_{D}$ accounts for the isochoric rearrangement due to local remodeling of living tissues. Therefore, the volumetric variation due to active growth $J_g=\det(\mathbf{F}_g)$ is only governed by the volumetric growth rate $\gamma$
\begin{equation}\label{eq:Jg_evol}
	\frac{d J_g}{dt} = \gamma J_g.
\end{equation}

Now we consider mass conservation for a growing region of tissue undergoing the transformation $\mathcal{B}_0\rightarrow\mathcal{B}_g\rightarrow\mathcal{B}_t$. By the decomposition of geometric deformation gradient, we have the multiplicative relation $J=\det(\mathbf{F})=J_eJ_g$, where $J_e=\det(\mathbf{F}_e)$ is the volumetric variation of the elastic deformation. Therefore, by change of variable, we have the mass of tissues within the region $\Omega_t\subset\mathcal{B}_t$ in the current deformed state 
\begin{equation}\label{eq:mass_conserv_changeofvariable}
\int_{\Omega_t}\rho(\mathbf{x},t) dV_t= \int_{\Omega_0} \rho(\mathbf{X},t)JdV_0=\int_{\Omega_0} \rho(\mathbf{X},t) J_eJ_g dV_0,
\end{equation}
where $\rho(\mathbf{X},t)=\rho(\mathbf{x}(\mathbf{X},t),t)$ is the local density attached to the material point initially located at $\mathbf{X}\in\Omega_0$ in the reference state $\mathcal{B}_0$. On the other hand, according the decomposition of deformation, we have assumed that the active growth induces mass change only by a volumetric change without changing the density. In contrast, the elastic deformation does not change the mass by inducing both volumetric and density changes. Therefore, mass changes in the current deformed state are only attributed to volumetric variations during the active growth stage
\begin{equation}\label{eq:mass_conserv_int}
	\int_{\Omega_t}\rho dV_t= \int_{\Omega_0} \rho_0 J_gdV_0,
\end{equation}
where $\rho_0(\mathbf{X})=\rho(\mathbf{x}(\mathbf{X},t),0)$ is the initial stress-free density, which is assumed to be constant. By the arbitrariness of the volume elements $\Omega_0$ and $\Omega_t$, we can obtain the relation from Eqs. \eqref{eq:mass_conserv_changeofvariable} and \eqref{eq:mass_conserv_int}, $\rho J_e = \rho_0\equiv const.$.
Furthermore, we can obtain the rate of change of total mass as
\begin{equation}\label{eq:mass_rate_balance}
	\frac{d}{dt}\int_{\Omega_t} \rho \ dV_t=\int_{\Omega_0} \frac{dJ_g}{dt}\rho_0dV_0=\int_{\Omega_t} \rho\gamma \ dV_t,
\end{equation}
where we have used the relations $dJ_g/dt=\gamma J_g$ and $\rho_0 J_g dV_0=\rho J_e J_g dV_0=\rho dV_t$ in the last equality. 
By the Reynolds transport theorem and the arbitrariness of $\Omega_t$, we then obtain the local mass conservation equation with a source
\begin{equation}\label{eq:mass_conserv}
	\frac{d\rho}{dt} =\frac{\partial \rho}{\partial t}+\mathbf{v}\cdot\nabla\rho= \rho(\gamma - \nabla\cdot \mathbf{v}).
\end{equation}

\begin{remark} 
In the case of an incompressible material, the local density remains constant $\rho\equiv\rho_0$ and hence we have the constraints for incompressibility
\begin{equation}
\nabla\cdot\mathbf{v}=\gamma\label{eq:incomp_cond} \quad \text{and} \quad J_e = 1.
\end{equation}
\end{remark}

\subsection{Dynamics of Elastic Deformation Tensor}
By the relation $\mathbf{F}_e=\mathbf{F} \mathbf{F}_g^{-1}$ and Eqs.~(\ref{eq:F_evol}) and (\ref{eq:Fg_evol}), we obtain the dynamics of the elastic deformation tensor $\mathbf{F}_e$
\begin{equation}\label{eq:Fe_evol}
\frac{d \mathbf{F}_e}{dt} = (\nabla\mathbf{v}-\boldsymbol{\Gamma})\mathbf{F}_e = (\nabla\mathbf{v}-\frac{\gamma}{d}\mathbf{I})\mathbf{F}_e-\boldsymbol{\Gamma}_{D}\mathbf{F}_e,
\end{equation}
where the Eulerian growth rate tensor $\boldsymbol{\Gamma}=\mathbf{F}_e\tilde{\boldsymbol\Gamma}\mathbf{F}_e^{-1}$ is the push-forward of the growth rate tensor from $\mathcal{B}_g$ to the current configuration $\mathcal{B}_t$ and $\boldsymbol{\Gamma}_{D}=\mathbf{F}_e\tilde{\boldsymbol\Gamma}_{D}\mathbf{F}_e^{-1}$ is the deviotoric part with isochoric rearragement in the Eulerian frame. Then the evolution of elastic volumetric variation follows
\begin{equation}\label{eq:Je_evol}
\frac{dJ_e}{dt}=J_e(\nabla\cdot\mathbf{v}-\gamma),
\end{equation}
which is consistent with the mass conservation with source in Eq.~(\ref{eq:mass_conserv}) via the relation $d(\rho J_e)/dt=0$ (derived from $\rho J_e = \rho_0 \equiv const.$). Thus, Eq.(\ref{eq:Fe_evol}) implies mass conservation given the decomposition Eq.(\ref{decomposition}). We note Eq.(\ref{eq:Fe_evol}) also implies the formula for the frame invariant evolution of the Finger elastic deformation tensor $\mathbf{B}_e=\mathbf{F}_e\mathbf{F}_e^\text{T}$ (see SM. ~\ref{FrameInvariance}):
\begin{equation}\label{eq:be_evol}
\frac{d\mathbf{B}_e}{dt} - \nabla \mathbf{v}\mathbf{B}_e-\mathbf{B}_e\nabla \mathbf{v}^\text{T}=-\frac{2}{d}{\gamma}\mathbf{B}_e- (\boldsymbol{\Gamma}_{D}\mathbf{B}_e+\mathbf{B}_e\boldsymbol{\Gamma}_{D}).
\end{equation}

\subsection{Compressible neo-Hookean Elasticity}

We next incorporate a constitutive law for the elastic stresses. We assume that the elastic energy $W$ can be defined solely in terms of the elastic deformation $\mathbf{F}_e$ and neglect any potential dependence on  $\mathbf{F}_g$ and $\mathbf{F}$. That is, $W=W(\mathbf{F}_e)$. Note that the tensors $\mathbf{F}_g$ and $\mathbf{F}$ can be reconstructed explicitly by solving Eqs. (\ref{eq:Fg_evol}) and (\ref{eq:F_evol}), respectively, without appealing to $\mathbf{F}_e$.

Considering the neo-Hookean elasticity model for isotropic, compressible tissues, the strain energy density function (per unit initial undeformed volume) in $d$-dimensions ($d=2,3$) is given by 
\begin{equation}\label{eq:W_elas}
W(\mathbf{F}_e)=\frac{1}{2}\mu(\bar{I}_1-d)+\frac{1}{2}K(J_e-1)^2,
\end{equation}
where $\mu$ and $K$ are, respectively, the shear and bulk modulus, and $\bar{I}_1=J_e^{-2/d}\tr(\mathbf{F}_e\mathbf{F}_e^\text{T})(=J_e^{-2/d}\tr(\mathbf{F}_e^\text{T}\mathbf{F}_e))$ is the first invariant of the isochoric part of the Finger elastic deformation tensor (Cauchy-Green elastic deformation tensor).

The Cauchy stress tensor for the compressible tissue with growth and rearrangement is then given by 
\begin{align} \label{stress}
\boldsymbol\sigma = {J_e}^{-1}\frac{\partial W}{\partial \mathbf{F}_e} \mathbf{F}_e^{\text T} = \mu J_e^{-\frac{2+d}{d}}\Big(\mathbf{F}_e\mathbf{F}_e^{\text{T}}-\frac{I_1}{d}\mathbf{I}\Big)+K(J_e-1)\mathbf{I},
\end{align}
where $I_1=\tr(\mathbf{F}_e\mathbf{F}_e^\text{T})$. The stress can be decomposed as $\boldsymbol{\sigma}=\boldsymbol{\sigma}_D+\boldsymbol{\sigma}_p$, the sum of the deviatoric part $\boldsymbol{\sigma}_D$ (proportional to $\mu$) and the isotropic part $\boldsymbol{\sigma}_p$ (proportional to $K$). The normal stress average is defined as ${\sigma}_{\text{N}} := \frac{\tr(\boldsymbol{\sigma})}{d}$. 

\begin{remark} [Incompressible Cauchy stress tensor]
We can also recover the incompressible neo-Hookean elastic stress as follows:
\begin{equation}
\boldsymbol\sigma = \mu \mathbf{F}_e\mathbf{F}_e^{\text{T}}-p\mathbf{I}.
\end{equation}
Compare to the compressible version, the pressure $p$ here plays the role of the Lagrangian multiplier for the incompressibility constraints, which also contains the term $\mu\frac{{I}_1}{d}\mathbf{I}$ appearing in the compressible model. In particular, $p =- {\sigma}_{\text{N}}+\mu\frac{{I}_1}{d}\mathbf{I}$.
\end{remark}

\section{Thermodynamically Consistent Modeling of Chemomechanical Feedback on Growth}

Next, we investigate the constraints on constitutive relations for the velocity $\mathbf{v}$, the volumetric growth rate $\gamma$ and the isochoric rearrangement rate tensor $\boldsymbol{\Gamma}_D$ that arise from requiring thermodynamic consistency \cite{gurtin2010mechanics} of the chemomechanical system.

\subsection{Elastic Energy Dissipation with Rearrangement}\label{subsec:Ew_decay}
Consider the total strain (elastic) energy of tissues stored within the deformed tumor region $\mathcal{B}_t$
\begin{equation}
E_w(t)=\int_{\mathcal{B}_t} J_e^{-1}W(\mathbf{F}_e) dV_t,
\end{equation}
where $J_e^{-1}W(\mathbf{F}_e)$ represents the strain energy density per deformed volume in the Eulerian frame. The rate of change of $E_w(t)$ is 
\begin{align}
\frac{d E_w}{dt} &= \frac{d}{dt}\int_{\mathcal{B}_t} J_e^{-1}W(\mathbf{F}_e)dV_t = \int_{\mathcal{B}_t} \Big(\frac{d}{dt}(J_e^{-1}W) + J_e^{-1}W \nabla \cdot \mathbf{v} \Big) dV_t, \nonumber  \\  
&= \int_{\mathcal{B}_t} \Big(J_e^{-1}\frac{\partial W}{\partial \mathbf{F}_e}:\frac{d\mathbf{F}_e}{dt} -J_e^{-2}W\frac{dJ_e}{dt} + J_e^{-1}W \nabla \cdot \mathbf{v} \Big) dV_t,  \nonumber  \\ 
&= \int_{\mathcal{B}_t} \Big(\boldsymbol{\sigma}: \nabla \mathbf{v} - \boldsymbol{\sigma}:\frac{\gamma}{d}\mathbf{I} +  J_e^{-1}W\gamma\Big) dV_t  -\int_{\mathcal{B}_t} \boldsymbol{\sigma}:\boldsymbol{\Gamma}_{D} \ dV_t, \label{eq:Ewdynamic}
\end{align}
where we have used the general formula for the stress $\boldsymbol{\sigma}$ in the first equality of Eq.~\eqref{stress}, and the dynamics of $\mathbf{F}_e$ and $J_e$ in Eqs.~\eqref{eq:Fe_evol} and \eqref{eq:Je_evol} in the last equality. The three terms in the first integral represent respectively the stress power due to local relative deformation, the stress power due to isotropic volumetric growth, and the addition (loss) of elastic energy due to volumetric growth (loss). 

Using the relation
$\boldsymbol{\sigma}:\nabla \mathbf{v} = \nabla\cdot(\boldsymbol{\sigma}\cdot\mathbf{v}) - (\nabla\cdot \boldsymbol{\sigma})\cdot\mathbf{v}$ and the divergence theorem, we can rewrite the stress power due to deformation as
\begin{equation}\label{eq:oldenergy}
\int_{\mathcal{B}_t} \boldsymbol{\sigma}:\nabla \mathbf{v} dV_t = -\int_{\mathcal{B}_t} (\nabla\cdot \boldsymbol{\sigma})\cdot\mathbf{v} dV_t + \int_{\partial \mathcal{B}_t} \mathbf{v}\cdot (\boldsymbol{\sigma}\cdot \mathbf{n}) dS_t,
\end{equation}
where $\mathbf{n}$ is the outward unit normal of the boundary $\partial \mathcal{B}_t$. 

Assuming negligible body and inertial forces and elastic equilibrium:
\begin{equation}\label{eq:mech_equil}
\nabla \cdot \boldsymbol{\sigma} = \mathbf 0,~~\rm{in}~~\mathcal{B}_t,
\end{equation}
together with boundary conditions
\begin{align}\label{eq:free_bc}
\boldsymbol{\sigma}\cdot \mathbf{n} &= \mathbf{F}_{ext}, \quad \text{at a moving part of the boundary $\partial\mathcal{B}_t$}\backslash\Sigma_C,\\
\mathbf{v} &= \mathbf0, \quad \text{at a fixed part of the boundary $\Sigma_C$}, \label{eq:fix_bc}
\end{align}
we obtain
$$
\int_{\mathcal{B}_t} \boldsymbol{\sigma}:\nabla \mathbf{v} dV_t=\int_{\partial \mathcal{B}_t\backslash\Sigma_C} \mathbf{v}\cdot \mathbf{F}_{ext}~ dS_t.
$$
%
In the particular case where $\mathbf{F}_{ext}=\mathbf{0}$, the traction-free boundary condition, this yields $\int_{\mathcal{B}_t} \boldsymbol{\sigma}:\nabla \mathbf{v}~ dV_t=\mathbf{0}$. Note that the velocity $\mathbf{v}$ is implicitly determined by the mechanical equilibrium \eqref{eq:mech_equil} and the boundary conditions \eqref{eq:free_bc}, e.g., by taking the material derivative (see Sec. \ref{num methods}). 
 
The second integral in Eq. \eqref{eq:Ewdynamic} represents the stress power due to the deviatoric growth that describes the effect of tissue rearrangement. Since $\boldsymbol{\sigma}:\boldsymbol{\Gamma}_D = \boldsymbol{\sigma}_D :\boldsymbol{\Gamma}_D=\tr(\boldsymbol{\sigma}_D\boldsymbol{\Gamma}_D)$, due to the fact $\tr(\boldsymbol{\Gamma}_D)=0$, this term is dissipative if
$\boldsymbol{\sigma}_D:\boldsymbol{\Gamma}_D\geq 0$.
%
%
For the compressible neo-Hookean elastic model in Eq. \eqref{eq:W_elas}, one possible choice of $\boldsymbol{\Gamma}_D$ that satisfies the energy-dissipation condition and traceless condition is
\begin{equation}\label{eq:gamma_D}
\boldsymbol{\Gamma}_D:= \beta\Big((\mathbf{F}_e\mathbf{F}_e^{\text{T}})-\frac{1}{d}\tr\Big((\mathbf{F}_e\mathbf{F}_e^{\text{T}})\Big)\mathbf{I}\Big),
\end{equation}
where $\beta$ is the rearrangement rate. With this choice, the growth anisotropy that is induced by the deviatoric stress tensor $\boldsymbol{\sigma}_D$ induces, in turn, the relaxation of the stress anisotropy from tissue rearrangement (e.g., recall Eq.~\eqref{eq:be_evol}). 
The remaining terms on the right hand side of Eq. (\ref{eq:Ewdynamic}) can either generate or dissipate energy (when only considering the elastic energy). These terms will be 
will be discussed together with the chemical energy in the next subsection, and will be seen to contribute to chemomechanical feedback on the volumetric growth in an energy-dissipative manner. 

\subsection{Chemomechanical Regulation}
Next we take into account the chemomechanical regulation of the volumetric growth such that the volumetric growth rate $\gamma$ depends on both the reaction-diffusion process of biochemical growth factors (e.g., nutrients) and the mechanical factors (e.g., strains and/or stresses). 
We assume that $c(\mathbf{x},t)$ represents the concentration (per unit mass) of the growth factor, its dynamics follows a convective reaction-diffusion process 
\begin{equation}
\frac{d (\rho c)}{dt} +(\rho c)\nabla\cdot\mathbf{v}=-\gamma_c \rho c+\nabla\cdot (D\rho \nabla c),
\end{equation}
where $\gamma_c>0$ is a constant uptake rate of growth factor and $D := L^2 > 0$ is the diffusion coefficient ($L$ is the diffusion length). We assume that $c=c_0$ at the boundary $\partial\mathcal{B}_t$, where $c_0$ is taken to be uniform. 
Using the mass conservation equation \eqref{eq:mass_conserv}, the above equation and boundary condition can be rewritten as
\begin{equation}\label{eq:c_evol}
\begin{dcases}
&\rho\frac{d c}{dt} =-(\gamma_c+\gamma) \rho c+\nabla\cdot (D \rho \nabla c),~~\text{in}~~\mathcal{B}_t\\
&c = c_0 \quad \text{at the external boundary $\partial\mathcal{B}_t$}.
\end{dcases}
\end{equation}
We introduce the following general form of the biochemical energy:
%
\begin{equation}
E_c=\int_{\mathcal{B}_t}\rho\mathcal{E}_c(c(x,t)) \ dV_t.
\end{equation}
The rate of change of the chemical energy is 
\begin{align}
\frac{d E_c}{dt}  
&=\int_{\mathcal{B}_t} \frac{d\rho}{dt}\mathcal{E}_c+ \mathcal{E}'_c\rho\frac{d c}{dt}+\rho \mathcal{E}_c (\nabla\cdot\mathbf{v}) dV_t \nonumber =\int_{\mathcal{B}_t} \rho \Big(\gamma \mathcal{E}_c-(\gamma_c+\gamma)\mathcal{E}'_cc\Big)+\mathcal{E}'_c\nabla\cdot (D \rho \nabla c) dV_t \nonumber \\ 
&= -\int_{\mathcal{B}_t} \rho\Big(\gamma(\mathcal{E}'_cc-\mathcal{E}_c)+\gamma_c\mathcal{E}'_cc +D\rho \mathcal{E}''_c|\nabla c|^2 \Big)~dV_t+ \int_{\partial\mathcal{B}_t} D\rho \mathcal{E}'_c (\nabla c\cdot \mathbf{n})dS_t.
%
%
%
%
\end{align}
where we have used the conservation of mass \eqref{eq:mass_conserv} and the dynamics of nutrient \eqref{eq:c_evol} in the second equality, and the divergence theorem in the last equality. The primes, $\mathcal{E}'_c$, $\mathcal{E}''_c$ denote the first and second derivatives with respect to $c$. The first term in the first integral couples the growth rate $\gamma$ with changes in the chemical energy while the remaining terms describe the change in energy due to uptake and diffusion of nutrients. The boundary integral describes the external energy influx.

We next define the total energy to be the sum of the elastic and biochemical energies:
\begin{equation}
E(t)=E_w(t)+E_c(t).
\label{total energy}
\end{equation}
%
%
%
%
%
%
%
The rate of change of the total energy is given by
\begin{align}
\frac{d E}{dt} = Q -\int_{\Omega_t} \Big(\boldsymbol{\sigma}_D:\boldsymbol{\Gamma}_D+\rho (\gamma_c\mathcal{E}'_cc +D\rho \mathcal{E}''_c|\nabla c|^2) \Big)dV_t +\int_{\partial\Omega_t} D\rho \mathcal{E}'_c (\nabla c\cdot \mathbf{n})dS_t.
\label{time deriv E}
\end{align}
If we assume that  $\mathcal{E}'_c\geq0$ and $\mathcal{E}''_c\geq0$, then the second term on the right hand side of Eq. (\ref{time deriv E}) is non-positive since we already assumed that $\boldsymbol{\sigma}_D:\boldsymbol{\Gamma}_D\ge 0$ in the previous section. The term $Q$ is defined as:
\begin{equation}
Q = -\int_{\Omega_t}\gamma\Big(\rho (\mathcal{E}'_cc-\mathcal{E}_c)  + (\tr(\boldsymbol{\sigma})/d-J_e^{-1}W\big)\Big) dV_t,
\label{Q equation}
\end{equation}
which couples the biochemical and elastic energies with growth. 
If we assume that the growth rate $\gamma$ is a non-negative (non-positive) function of the term in parenthesis in Eq. (\ref{Q equation}), then the coupling term $Q$ is non-positive (non-negative). The simplest choice is to take the growth rate to be 
\begin{equation}\label{eq:gamma_feedback}
\gamma = \eta \Big(\rho (\mathcal{E}'_cc-\mathcal{E}_c) + \big(\sigma_N-J_e^{-1}W\big)\Big),
\end{equation}
where $\eta$ is a positive  scaling factor that might in general depend on the nutrient concentration $c$ and the stress $\boldsymbol{\sigma}$.  This provides a natural form for the chemomechanical feedback on the growth rate $\gamma$ that depends on both the nutrient concentration and mechanical stress.

\begin{remark}
A similar approach would also give the chemomechanical feedback on the growth rate under incompressible assumption 
\begin{equation}
\gamma = \eta\Big(\mathcal{E}'_cc-\mathcal{E}_c + \big(\sigma_N-W\big)\Big).
\end{equation}
Notice that this formula involves the elastic stress and pressure since in the incompressible case, $\sigma_N=-p+\frac{\mu}{d}\tr({\mathbf{B}_e})$. 
\end{remark} 

\subsection{Comparison with other biomechanical models} \label{subsec:other_growth_models}
We have proposed a general continuum framework for the chemomechanical regulation of the tissue growth and mechanics based on the theory of finite elasticity and growth. 
This framework is essentially different from many previous nonlinear finite growth models \cite{Goriely-2005-prl,Li-2011-jmps,oltean2018apoptosis} where the growth tensor is prescribed as either piecewise constant or linear functions of radius while its dynamical process is not depicted. A constant growth rate will lead to unphysical exponential growth of tissue and infinite accumulation of elastic stress, which emphasizes the need for growth regulation and stress relaxation \cite{ambrosi2002mechanics}. In our approach, we specifically describe the evolution of the growth tensor subject to biochemomechanical regulation. 

Many previous tissue growth regulation models can be easily recovered by our framework under some appropriate assumptions on the growth stretch tensor and its growth rate tensor, some of which will be mentioned below. Budday {\it et al} \cite{budday2015size} studied the regulation of growth in mammalian brains, where the brain was modeled as a bilayered system of growing outer gray matter layer covering an inner white matter core. In particular, they considered isotropic growth for the white matter core that is activated by an excess elastic stretch induced by the active growth of outer gray matter layer: 
\begin{equation}
    \mathbf{F}_g = J_g^{\frac{1}{3}} \mathbf{I},\quad \frac{d J_g}{dt} = G_w (J_e - 1),
\end{equation}
where $G_w$ is the axon elongation rate that reflects the mechanical feedback on the gray matter growth. Our chemomechanical regulation model reduces to this model when assuming there is no tissue rearrangement $\boldsymbol{\Gamma}_D = \mathbf{0}$ in Eq. \eqref{eq:Gamma_decomp}, and there is a balance between chemical and elastic energy $\rho (\mathcal{E}'_cc-\mathcal{E}_c) =J_e^{-1}W$ for the regulation on the volumetric growth rate $\gamma$ in \eqref{eq:gamma_feedback}, which implies the relation $G_w = K\eta J_g $. 

Garcia {\it et al} \cite{garcia2017contraction} considered a stress-dependent isotropic growth rate tensor $\boldsymbol{\Gamma}$ for the growth of the brain tube of an early chicken embryo by
\begin{equation}
     \frac{d \mathbf{F}_g}{dt} = \boldsymbol\Gamma \mathbf{F}_g,\quad \boldsymbol\Gamma =  (g_0+g_\sigma \overline{\sigma}) \mathbf{I},
\end{equation}
where $g_0 \geq 0$ represents the baseline growth rate, $g_\sigma \geq 0$ is a coefficient for stress-dependent growth, and $\overline{\sigma}$ denotes the non-dimensionalized average in-plane stress of the lateral surface. This formulation aligns with our model when assuming no tissue arrangement $\boldsymbol{\Gamma}_D = \mathbf{0}$ and the feedback on volumetric growth rate is $\gamma := g_0 + g_\sigma \overline{\sigma}$. Note that without the feedback ($g_\sigma=0$) this will lead exponential growth. Although this feedback on $\gamma$ shares a similar idea with our model in that the growth rate is governed by both chemical factor (with $g_0$) and mechanical factor (with $\overline{\sigma}$), they did not consider the energy change due to the feedback.

While the above two models only assumed isotropic growth rate regulated mechanically by the average elastic stretch or stress and without chemical regulation, Arajo and McElwain \cite{araujo2004linear} connected the growth rate with the nutrient concentration and the stress anisotropy for incompressible elastic avascular tumor by
\begin{align}\label{eq:Arajo_gamma}
  \boldsymbol{\Gamma} &:= q~\mathbf{diag}(\eta_r,\eta_\theta,\eta_\theta) - h \ \mathbf{diag}(\zeta_r,\zeta_\theta,\zeta_\theta)\nonumber \\
  &= \frac{1}{3}(q-h)\mathbf{I} + \Big(q~\big(\mathbf{diag}(\eta_r,\eta_\theta,\eta_\theta)-\frac{1}{3}\mathbf{I}\big)- h~ \big(\mathbf{diag}(\zeta_r,\zeta_\theta,\zeta_\theta)-\frac{1}{3}\mathbf{I}\big)\Big),
\end{align}
where $q$ and $h$ respectively present the growth rates related to cell proliferation and apoptosis, and $\eta$'s are proliferation-strain multipliers such that $\eta_r + 2\eta_\theta = 1$ and $\zeta$'s are apoptosis strain multipliers such that $\zeta_r+2\zeta_\theta = 1$. In particular, $q$ and $h$ are directly related to the equilibrium nutrient concentration, and $\eta$'s and $\zeta$'s are prescribed functions of the stress anisotropy $(\sigma_{rr}-\sigma_{\theta\theta})$. If we decompose the Eulerian growth rate tensor as in \eqref{eq:Arajo_gamma}, this chemomechanical feedback can be also captured by our model by assuming the volumetric growth rate $\gamma:= q-h$ and the deviatoric growth rate $\boldsymbol{\Gamma}_D:= q~\big(\mathbf{diag}(\eta_r,\eta_\theta,\eta_\theta)-\frac{1}{3}\mathbf{I}\big)- h~ \big(\mathbf{diag}(\zeta_r,\zeta_\theta,\zeta_\theta)-\frac{1}{3}\mathbf{I}\big)$. However, Arajuo and McElwain did not consider the thermodynamics of the feedback either, and more importantly, they did not have the freedom to control the anisotropic growth remodeling without volumetric growth.  

In \cite{garcke2022,garcke2024}, a thermodynamically consistent viscoelastic model of tumor growth was developed. In contrast to our approach, the elastic relaxation in \cite{garcke2022,garcke2024} can induce local mass changes. Further, regulation of growth by biochemical and elastic forces used a Hill function form involving the Frobenious norm of the stress tensor \cite{garcke2021}. Hill-type forms of the biomechanical regulation of growth laws were also considered in \cite{ambrosimollica2004,ambrosi2012,ciarletta2013,ambrosi2017solid, yan2021stress}. A history-dependent approach for biomechanical regulation of growth laws was recently considered in \cite{walker2023}.

In our previous work in \cite{yan2021stress}  we also incorporated stress relaxation and mechanical feedback on growth and modeled the deformation tensors using an adaptive reference map. While the dynamics of the deformation tensors was frame invariant for spherical tumors, this was not the case in general. Further, unlike the approach we develop here, neither the stress relaxation nor the biomechanical feedback we considered were energy dissipative.

\subsection{Model Summary and Non-dimensionalization}\label{subsec:nondim}
From now on, we make the following choices for the energy density functions and feedback parameters: (i) the neo-Hookean elastic model \eqref{eq:W_elas} for the compressible tissue; (ii) a quadratic form for the chemical energy density $\mathcal{E}_c=\frac{k}{2}c^2$ with an energy coefficient $k$; (iii) the simple form for the deviatoric growth $\boldsymbol{\Gamma}_D$ as in Eq. \eqref{eq:gamma_D}; (iv) a linear dependence of the scaling factor in Eq. \eqref{eq:gamma_feedback} on the nutrient concentration $\eta(c,\boldsymbol{\sigma}):=\eta c$ (with $\eta>0$ being a constant), motivated by the fact that the proliferation rate $\gamma$ should be sensitive to the overall concentration of nutrient. 

In summary, our coupled minimal system consists of tissue growth and mechanics, and their interactions with a biochemical reaction diffusion process:
\begin{align*}
\begin{dcases}
\text{Mechanical Equilibrium: } 
\nabla \cdot \boldsymbol{\sigma} = \mathbf 0, \quad 
\boldsymbol\sigma = \mu J_e^{-\frac{2+d}{d}}\Big(\mathbf{F}_e\mathbf{F}_e^{\text{T}}-\frac{\tr(\mathbf{F}_e\mathbf{F}_e^{T})}{d}\mathbf{I}\Big)+K(J_e-1)\mathbf{I}, \\
\text{Elastic Deformation Dynamics: }
\frac{d \mathbf{F}_e}{dt} = (\nabla\mathbf{v}-\frac{\gamma}{d}\mathbf{I})\mathbf{F}_e-\beta\Big((\mathbf{F}_e\mathbf{F}_e^{\text{T}})-\frac{1}{d}\tr\Big((\mathbf{F}_e\mathbf{F}_e^{\text{T}})\Big)\mathbf{I}\Big)\mathbf{F}_e, \\
\text{Biochemical Reaction-Diffusion Process: }
\rho\frac{d c}{dt} =-(\gamma_c+\gamma) \rho c+\nabla\cdot (D \rho \nabla c),\\
\text{Biochemomechanical Feedback on Growth: }
\gamma =\eta c~ \Big(\frac{1}{2}k \rho c^2 + \big(K(J_e-1)-J_e^{-1}W\big)\Big), \\
\end{dcases}
\end{align*}
subject to the initial conditions 
\begin{equation*}
\mathbf{F}_e(\mathbf{x},0) = \mathbf I, \quad
c(\mathbf{x},0) = c_0,
\end{equation*}
and mixed boundary conditions
\begin{align*}
\begin{dcases}
c(\mathbf{x},t)=c_0, \quad \text{at the external boundary $\partial \mathcal{B}_t$}, \\
\boldsymbol{\sigma}\cdot \mathbf{n} = \mathbf{F}_{ext}, \quad  \text{at a moving part of the boundary $\Sigma_t=\partial\mathcal{B}_t\backslash\Sigma_C$},\\
\mathbf{v} = \mathbf0, \quad \text{at a fixed part of the boundary $\Sigma_C$}.
\end{dcases}
\end{align*}
Here we have assumed a stress-free initial condition for the elastic deformation tensor and uniform initial condition for nutrient concentration. Note that the velocity field $\mathbf{v}$ is implicitly determined by the mechanical equilibrium condition and boundary conditions (see Sec. \ref{num methods}). Once given the velocity field, the elastic deformation tensor $\mathbf{F}_e$ and the nutrient concentration field $c$ can be updated. Then, the growth rate $\gamma$ is updated correspondingly. Furthermore, we can also reconstruct and track other interesting quantities such as the density $\rho$, the deformation gradient $\mathbf{F}$ and growth tensor $\mathbf{F}_g$ accordingly. 

We nondimensionalize the equations using a length scale $l=1~\mu m$ and time scale, $\tau=1~\rm{day}$ (see SM. \ref{nondimparam} for details). The resulting system is equivalent to the above equations where we set $\mu = 1$, and $c_0 = 1$.

\begin{remark}
  The minimal system for incompressible case after non-dimensionalization is the following equations:

\begin{align*}
\begin{dcases}
\text{Mechanical Equilibrium: } 
\nabla \cdot \boldsymbol{\sigma} = \mathbf 0,\quad \boldsymbol\sigma = \mathbf{F}_e\mathbf{F}_e^T - p\mathbf{I},\\
\text{Elastic Deformation Dynamic: }
\frac{d \mathbf{F}_e}{dt} = (\nabla\mathbf{v}-\frac{\gamma}{d}\mathbf{I})\mathbf{F}_e-\beta\Big((\mathbf{F}_e\mathbf{F}_e^{\text{T}})-\frac{1}{d}\tr\Big((\mathbf{F}_e\mathbf{F}_e^{\text{T}})\Big)\mathbf{I}\Big)\mathbf{F}_e, \\
\text{Chemical Reaction-Diffusion Process: }
\frac{d c}{dt} =-(\gamma_c+\gamma)c+\nabla\cdot (D \nabla c), \\
\text{Biochemomechanical Feedback on Growth: }
\gamma = \eta c~\Big(\frac{1}{2}k c^2 - \big( p-\frac{\mu \tr(\mathbf{F}_e\mathbf{F}_e^T)}{d}+W\big)\Big), \\
\text{Incompressibility condition: }
\nabla\cdot\mathbf{v}=\gamma.
\end{dcases}
\end{align*}
The system's initial and boundary conditions are equivalent to the compressible system.

\end{remark}

\section{Model Reduction to Radial Symmetry and Numerical Method}
To demonstrate the effectiveness of the proposed framework, we will employ our model to study the growth of spheroidal tissues in the next section. Here, we provide the reduced model for spherical geometry and the numerical method for simulation. 
\subsection{Systems with radial symmetry}
Consider the system in spherical coordinates with radial symmetry in three dimensions. We have
\begin{align}
&\mathbf{F}_e = \textbf{diag}({f_e}_r,{f_e}_\theta,{f_e}_\theta), \quad J_e = {f_e}_r{f_e}_\theta^2,\\
&\mathbf{v} = (v,0,0)^{\text{T}}, \quad
\nabla \mathbf{v} = \textbf{diag}(v_r, \frac{v}{r},\frac{v}{r}),\\
&\boldsymbol\sigma = \textbf{diag}(\sigma_{rr},\sigma_{\theta\theta},\sigma_{\theta\theta}).
\end{align}
We want to solve the following system for $\{{f_e}_r,{f_e}_\theta,v,R\}$: 
\begin{equation}
\begin{dcases}
&\frac{\partial {{f_e}_r}}{\partial t}+v\frac{\partial {{f_e}_r}}{\partial r}=\Big(v_r-\frac{\gamma}{3}-\frac{2}{3}\beta( {{f_e}_r}^2-{{f_e}_\theta}^2)\Big){{f_e}_r}\\   
&\frac{\partial {{f_e}_\theta}}{\partial t}+v\frac{\partial {{f_e}_\theta}}{\partial r}=\Big(\frac{v}{r}-\frac{\gamma}{3}-\frac{1}{3}\beta( {{f_e}_\theta}^2-{{f_e}_r}^2)\Big){{f_e}_\theta} \\
&\frac{\partial \sigma_{rr}}{\partial r}+\frac{2}{r}(\sigma_{rr}-\sigma_{\theta\theta})=0,\\
&\sigma_{rr}= \frac{2}{3} J_e^{-\frac{5}{3}}({f_e}_r^2-{f_e}_\theta^2)+K(J_e-1), \\ 
&\sigma_{\theta\theta}= \frac{1}{3} J_e^{-\frac{5}{3}}({f_e}_\theta^2-{f_e}_r^2)+K(J_e-1),\\
&\rho(\frac{\partial c}{\partial t} + v\frac{\partial c}{\partial r}) =-(\gamma_c+\gamma) \rho c+\frac{D}{r^2}\frac{\partial}{\partial r}(r^2\rho\frac{\partial c}{\partial r}) ,\\
\end{dcases}	
\end{equation}
subject to the initial and boundary conditions
\begin{equation} \label{eq:bc}
\begin{dcases}
{f_e}_r(r,0)={f_e}_\theta(r,0)=1, \quad R(0)=R_0, \quad \text{at $t=0$}\\
v(0,t)=0, \quad{f_e}_r(0,t)={f_e}_\theta(0,t)=1, \quad \text{at $r=0$}\\
\sigma_{rr}(R,t)=F_{ext},\quad \frac{dR}{dt} = v(R,t), \quad \text{at $r=R(t)$}.
\end{dcases}
\end{equation}
Assuming that the tissue is embedded in gel with relative rigidity $c_H$ (relative to the tumor shear modulus $\mu$), we have an analytic expression for the external traction $F_{ext}$ at the tissue-gel interface \cite{yan2021stress}:
\begin{equation}
	F_{ext}(t) = \frac{c_H}{2}(5-\frac{R_0(R_0^3+4R(t)^3)}{R(t)^4}).
 \label{chforceEQ}
\end{equation}

Solving the above system, which is a highly coupled nonlinear system with moving boundary conditions, is challenging. Following \cite{yan2021stress}, we can tackle this one-dimensional system by rescaling. We first introduce the change of variable $r'=r/R(t)$ such that the moving boundary problem of the original system is reduced to a problem in a fixed domain for $r'\in [0,1]$. With the change of variable, for any function $f(r,t)$, we have the partial derivatives 
\begin{align}
&\partial_{r} f(r,t) = \partial_{r'}f(r',t)/R(t),\\	
&\partial_{t} f(r,t) = \partial_{t}f(r',t)-\frac{r'\dot{R}}{R}\partial_{r'}f(r',t)
\end{align}
From now on, we will drop the primes for convenience (i.e., $r/R(t)\rightarrow r$). The system becomes

\begin{empheq}[left=\empheqlbrace]{align} 
	&\frac{\partial {{f_e}_r}}{\partial t}+\tilde{v}\frac{\partial {{f_e}_r}}{\partial r}=\Big(\frac{v_r}{R}-\frac{\gamma}{3}-\frac{2}{3}\beta( {{f_e}_r}^2-{{f_e}_\theta}^2)\Big){{f_e}_r}, \label{eq:fer_rescale}\\   
	&\frac{\partial {{f_e}_\theta}}{\partial t}+\tilde{v}\frac{\partial {{f_e}_\theta}}{\partial r}=\Big(\frac{v}{rR}-\frac{\gamma}{3}-\frac{1}{3}\beta( {{f_e}_\theta}^2-{{f_e}_r}^2)\Big){{f_e}_\theta}, \label{eq:fet_rescale} \\
	&\frac{\partial \sigma_{rr}}{\partial r}+\frac{2}{r}(\sigma_{rr}-\sigma_{\theta\theta})=0, \label{eq:force_balance_rescale}\\
	&\sigma_{rr}= \frac{2}{3} J_e^{-\frac{5}{3}}({f_e}_r^2-{f_e}_\theta^2)+K(J_e-1), \label{eq:str_rr_rescale}\\ 
	&\sigma_{\theta\theta}= \frac{1}{3} J_e^{-\frac{5}{3}}({f_e}_\theta^2-{f_e}_r^2)+K(J_e-1),\label{eq:str_tt_rescale}\\
 &\rho(\frac{\partial c}{\partial t} + \tilde{v}\frac{\partial c}{\partial r}) =-(\gamma_c+\gamma) \rho c+\frac{D}{r^2R^2}\frac{\partial}{\partial r}(r^2\rho\frac{\partial c}{\partial r}), \label{eq:nutrient}
\end{empheq}	
where $\tilde{v} = (v-r\dot{R})/R$ and $J_e = {f_e}_r{f_e}_\theta^2$. The system has the initial and boundary conditions
\begin{equation}\label{eq:bc_rescale}
	\begin{dcases}
		{f_e}_r(r,0)={f_e}_\theta(r,0)=1, \quad R(0)=R_0, \quad \text{at $t=0$}\\
		v(0,t)=0, \quad{f_e}_r(0,t)={f_e}_\theta(0,t)=1, \quad \text{at $r=0$}\\
		\sigma_{rr}(1,t)=F_{ext},\quad \frac{dR}{dt} = v(1,t), \quad \text{at $r=1$}.
	\end{dcases}
\end{equation}

\begin{remark}
In the incompressible case, the problem reduces to solving the following system for $\{{f_e}_r,{f_e}_\theta,v,p,R\}$: 
\begin{equation}
\begin{dcases}   
&\frac{\partial {{f_e}_\theta}}{\partial t}+\tilde{v}\frac{\partial {{f_e}_\theta}}{\partial r}=\Big(\frac{v}{rR}-\frac{\gamma}{3}-\frac{1}{3}\beta( {{f_e}_\theta}^2-\frac{1}{{{f_e}_\theta}^4})\Big){{f_e}_\theta} \\
& v = \frac{1}{r^2R^2}\int_{0}^{rR} \gamma s^2 ds, \\ 
&\frac{\partial \sigma_{rr}}{\partial r}+\frac{2}{rR}(\sigma_{rr}-\sigma_{\theta\theta})= \frac{1}{R}\frac{\partial p}{\partial r},\\
&\sigma_{rr}= 1/{f_e}_\theta^4, \quad \sigma_{\theta\theta}= {f_e}_\theta^2,\\
& \frac{\partial c}{\partial t}+\tilde{v}\frac{\partial c}{\partial r} =-(\gamma_c+\gamma)c+D  (\frac{1}{R^2}\frac{\partial^2 c}{\partial r^2}+\frac{2}{rR^2}\frac{\partial c}{\partial r}).\\
\end{dcases}	
\end{equation}

Notice that we have used the relation ${f_e}_r = {f_e}_\theta^{-2}$ since $J_e = 1$. Hence, we only need to compute the dynamics of either ${f_e}_r$ or ${f_e}_\theta$. In this paper, we opt to compute ${f_e}_\theta$ because it does not involve the gradient of velocity $v_r$. The system's initial and boundary conditions are equivalent to the compressible system.

\end{remark} 

\subsection{Numerical Methods}
\label{num methods}
The main challenge in solving the compressible system comes from the nonlinear coupling between the velocity field $v$ and the elastic strains ${f_e}_{r}$ and ${f_e}_{\theta}$, both explicitly by the evolution equations (\ref{eq:fer_rescale}-\ref{eq:fet_rescale}) and implicitly by the mechanical equilibrium equations (\ref{eq:force_balance_rescale}-\ref{eq:str_tt_rescale}) together with the boundary conditions. 
Instead of directly considering the mechanical equilibrium equation (\ref{eq:force_balance_rescale}), we consider a surrogate equation by taking its time derivative 
\begin{equation}
	\frac{d}{dt}(\nabla\cdot\boldsymbol{\sigma})+\tilde{\beta} \nabla\cdot\boldsymbol{\sigma}=0,
\end{equation}
where we have used the technique \cite{Jones2000} of adding an artificial damping term with $\tilde{\beta}$, which is chosen to be large enough to maintain the force balance condition $\nabla\cdot\boldsymbol{\sigma}=0$ at all times. 

The above equation in the radially symmetric case reduces to
\begin{align}
	&\Big(\frac{\partial}{\partial r}(\frac{\partial \sigma_{rr}}{\partial {f_e}_r})+\frac{2}{r}\frac{\partial \sigma_s}{\partial {f_e}_r}\Big)\frac{d {f_e}_r}{dt}+\Big(\frac{\partial}{\partial r}(\frac{\partial \sigma_{rr}}{\partial {f_e}_\theta})+\frac{2}{r}\frac{\partial \sigma_s}{\partial {f_e}_\theta}\Big)\frac{d {f_e}_\theta}{dt}+\frac{\partial \sigma_{rr}}{\partial {f_e}_r}\frac{\partial}{\partial r}(\frac{d {f_e}_r}{dt}) \nonumber\\
	&+\frac{\partial \sigma_{rr}}{\partial {f_e}_\theta}\frac{\partial}{\partial r}(\frac{d {f_e}_\theta}{dt})-\Big(\frac{\partial \sigma_{rr}}{\partial {f_e}_r}\frac{\partial {f_e}_r}{\partial r}+\frac{\partial \sigma_{rr}}{\partial {f_e}_\theta}\frac{\partial {f_e}_\theta}{\partial r}\Big)\frac{1}{R}\frac{\partial v}{\partial r}-\frac{2v}{r^2R}\sigma_s +\tilde{\beta}\Big(\frac{\partial \sigma_{rr}}{\partial r}+\frac{2\sigma_s}{r}\Big) \nonumber =0,
\end{align}
where $\sigma_s=\sigma_{rr}-\sigma_{\theta\theta}$. By combining the evolution equations (\ref{eq:fer_rescale}-\ref{eq:fet_rescale}) and the constitutive laws in equations (\ref{eq:str_rr_rescale})-(\ref{eq:str_tt_rescale}), we derive a comprehensive equation for updating the velocity in the bulk:
\begin{equation}\label{eq:solve_v_bulk}
a_1 v+a_2\frac{\partial v}{\partial r}+a_3\frac{\partial^2 v}{\partial r^2}+a_4=0, \quad \text{for $0<r<1$},
\end{equation}
We treat the external traction boundary condition in the same manner and add a damping term
\begin{equation}
\frac{d}{dt}(\sigma_{rr}-F_{ext})+\tilde{\beta}(\sigma_{rr}-F_{ext})=0, \quad \text{at $r=1$}, 
\end{equation}
which yields an equation for the velocity at the boundary 
\begin{equation}\label{eq:solve_v_boundary}
	a_5\frac{\partial v}{\partial r}+a_6\frac{v}{r}=a_7, \quad \text{for $r=1$},
\end{equation}
where the coefficients $a_k$ ($k=1,\ldots,7$) are  functions of $\{r,R,{f_e}_r,{f_e}_{\theta},\gamma\}$ and their first derivatives. 

Now the whole system reduces to solving Eqs. (\ref{eq:solve_v_bulk}) and (\ref{eq:solve_v_boundary} for the velocity field. We use a semi-implicit numerical method to decouple the system, where we solve firstly solve the velocity field by treating the coefficients $a$'s explicitly and then update the other variables $\{{f_e}_r,{f_e}_\theta,R\}$ with the obtained velocity. The numerical algorithm proceeds by iterating through the following three steps:
\begin{itemize}
\item[Step 1:] Given ${f_e}_r^n$, ${f_e}_\theta^n$ and $R^n$, compute the coefficients $a_k$ ($k=1,\ldots,7$) for the system of velocity (\ref{eq:solve_v_bulk}) and (\ref{eq:solve_v_boundary}). Any consistent finite difference approximation can be applied. Here we employ a five-point finite difference discretization for all the derivatives. This semi-implicit treatment allows us to derive a linear system for the velocity $v^n$, which we solve using MATLAB's integrated linear solver. 
\item[Step 2:] With $v^n$ obtained, we update ${f_e}_r^{n+1}$ and ${f_e}_\theta^{n+1}$ via their evolution equations (\ref{eq:fer_rescale}) and (\ref{eq:fet_rescale}) using an explicit second-order upwind scheme. Here the second-order method is used to ensure the first-order accuracy of the coefficients $a_k$ in Eq. \eqref{eq:solve_v_bulk} that involves $\partial_r{f_e}_r^{n+1}$ and $\partial_r{f_e}_\theta^{n+1}$.
\item[Step 3:] Update the boundary radius by $R^{n+1}=R^{n}+\Delta t v^n_N$.
\item[Step 4:] Compute $c^{n+1}$ by discretizing Eq. (\ref{eq:nutrient}) semi-implicitly and deriving a linear system as below
\begin{align}
\frac{c_i^{n+1}-c^n_i}{\Delta t} + \tilde{v}^{n+1} \frac{\partial c^n}{\partial r} = -(\gamma_c+\gamma^n_i) c^{n+1}_i + \frac{D}{r_i^2{R^{n+1}}^2}\frac{1}{\Delta r}\Big((r^2\rho\frac{\partial c}{\partial r})^{n+1}_{i+\frac{1}{2}}-(r^2\rho\frac{\partial c}{\partial r})^{n+1}_{i-\frac{1}{2}}\Big)
\end{align}
where the second term on the left hand side will be discretized explicitly by using the second order upwind scheme, and the righ hand side terms are defined as $(r^2\rho\frac{\partial c}{\partial r})_{i+\frac{1}{2}}=\frac{1}{2\Delta r}(r_i^2\rho_i+r_{i+1}^2\rho_{i+1})(c_{i+1}-c_i)$.
\end{itemize}

\begin{remark}
For the incompressible case, the process is more straightforward as we have a simple equation for velocity $v$ (Eq. \ref{eq:incomp_cond}), and we decouple the evolution of $f_e$ from the force balance equation. In the incompressible case, our numerical algorithm is:
\begin{itemize}
    \item[Step 1:] Given $fe_r^n$, $fe_\theta^n$. Calculate $\sigma_{rr}^n, \sigma_{\theta \theta}^n$.
    \item[Step 2:] Solve the coupled equations for the velocity $v^{n+1}$ and pressure ${p}^{n+1}$ : 
    \begin{equation}
        v^{n+1} = \frac{1}{r^2} \int_0^r \gamma(\sigma_{\theta\theta}^n,{p}^{n+1})s^2  ds ,\quad         \frac{\partial{p}^{n+1}}{\partial r}  =  \frac{\partial\sigma_{rr}^n}{\partial r} + \frac{2}{r}(\sigma_{rr}^n-\sigma_{\theta\theta}^n)
        \label{veloEQ}
    \end{equation}
    \begin{equation}
        \frac{\partial\sigma_{rr}^n}{\partial r} = -\frac{8}{3}(fe_\theta^n)^{-5} \frac{\partial fe_\theta^n}{\partial r}-\frac{4}{3}(fe_\theta^n) \frac{\partial fe_\theta^n}{\partial r}
    \end{equation}
    with boundary condition
    \begin{equation}
      {p}^{n+1}_{end} = (fe_{r,end}^n)^{2} + F_{ext}; 
    \end{equation}
  by differentiating Eq. \eqref{veloEQ} in $r$, using
   second order backward finite differences to approximate $v_{,r}$ and a second order forward difference for ${p}_{,r}$ and then construct a linear matrix to compute $v^{n+1},p^{n+1}$ in the radial direction. Update $\gamma^{n+1}$ with $p^{n+1}$. Then we check the residual of the linear matrix. If the residual is still larger than the tolerance (here we used $10^{-6}$), repeat the linear solver. If this requires repeating more than a specified number of times (we used 100), we apply "fsolve" function in Matlab using the newest $v^{n+1},p^{n+1}$ as the initial guess. 
    \item[Step 3:] Update the boundary radius by $R^{n+1}=R^{n}+\Delta t v_N$.
    \item[Step 4:] Given ${fe_r}^n,{fe_\theta}^n,v^n$. Solve for ${fe_r}^{n+1}$ and ${fe_\theta}^{n+1}$ explicitly using the upwind second order scheme. 
    \item[Step 5:]
     Then for radial position $i = 2$ to $N-1$, we have the implicit discretized formula for the nutrient equation as follows
\begin{equation}
\frac{c_i^{n+1}-c^n_i}{\Delta t} + \tilde{v} \frac{\partial c}{\partial r} = -(\gamma_c+\gamma^n_i) c^{n+1}_i+ D(\frac{c^{n+1}_{i+1}-2c^{n+1}_i+c^{n+1}_{i-1}}{(R^{n+1})^2(\Delta r)^2}+\frac{c_{i+1}^{n+1}-c^{n+1}_{i-1}}{(R^{n+1})^2r\Delta r})
\end{equation}
where the second term on the left hand side will be solved using the second order upwind scheme. The solution for $c^{n+1}$ is obtained by solving the linear system with the initial condition: $c^0_i = 1$ and the boundary conditions:$\frac{\partial c}{ \partial r} = 0$ at the origin ($i=1$) and $c^n_N = 1$  at the external boundary ($i=N$).
\end{itemize}
\end{remark}

\section{Results}
\subsection{Simulation of tumor growth and fitting to experimental data}
\begin{figure}
  \scalebox{0.62}[0.62]{\includegraphics{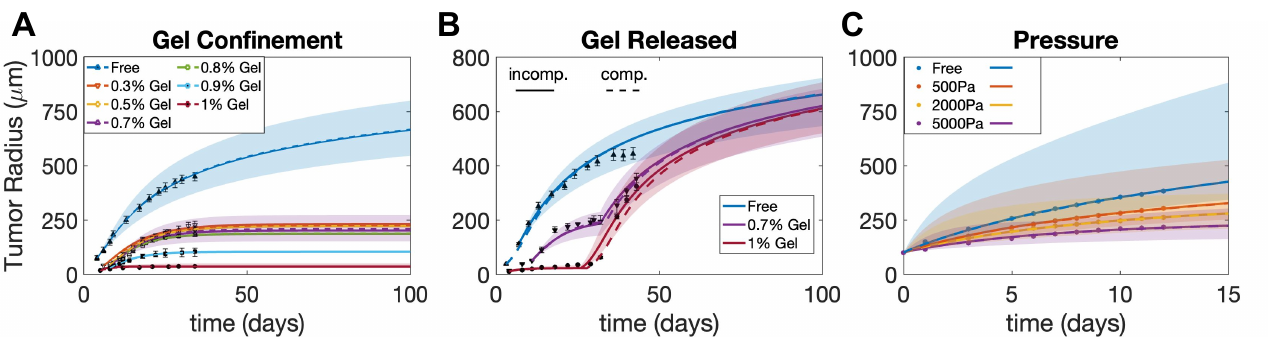}}
    \caption{A) Model fits (curves) to the data (symbols) of the tumor size evolution in free suspension, and in 0.3\%, 0.5\%, 0.7\%, 0.8\%, 0.9\% and 1.0\% agarose gels from Fig. 1A in \cite{helmlinger1997solid}.  B) Model fits (curves) to the data (symbols with error bars) when stresses are released by removing the gels in a narrow interval around the times reported in Fig. 1B in \cite{helmlinger1997solid}. C) Model fits (curves) of the data (symbols with errorbars) of the tumor size evolution in free static suspension, $500$ Pa, $2000$ Pa and $5000$ Pa from Fig 1 in \cite{Montel2021}. Bands indicate results that are within $10\%$ of the best fitting set of parameters. Solid curves: Incompressible model, dashed curves: Compressible model.} 
  \label{FittingRadius}
\end{figure} 
 
In this section, we investigate biochemomechanical regulation of tumor spheroid growth using our new model. It is well known that tumor cells respond to stresses through reducing proliferation, inducing apoptosis/necrosis, enhancing extracellular matrix (ECM) deposition/organization, and augmenting their invasive and metastatic potentials \cite{yu2011forcing}. At the tissue level, it is well documented  that different levels of external mechanical forces modulate the growth of tumor spheroids \cite{helmlinger1997solid,Montel2021,montel2012stress,delarue2014compressive,mascheroni2017evaluating}. One of the earliest experiments was performed by Helmlinger et al. \cite{helmlinger1997solid} where tumor spheroids were cultured in agarose gel environments with varying stiffness controlled by different concentrations of the agarose. The growth of tumor spheroids was notably inhibited when the agarose concentration was increased (as shown in Fig.~\ref{FittingRadius}A). Strikingly, when the gel was removed, the tumor spheroids resumed the growth at rates similar to their free-suspension counterparts (as shown in Fig.~\ref{FittingRadius}B). These growth behaviors make the experiment \cite{helmlinger1997solid} a suitable candidate to test and validate our proposed model. In addition, we also test our model using data from \cite{Montel2021} where tumor growth was inhibited by an applied osmotic pressure.

\subsubsection{Fitting with tumor growth experiments under gel confinement and external pressure}

To reproduce the tumor growth behaviors observed in \cite{helmlinger1997solid} with our model, we first find the appropriate model parameters (listed in Table \ref{table1} in SM. \ref{nondimparam}) by fitting with the experimental data of tumor growth in Fig.~\ref{FittingRadius}A. We choose the three sets of experimental data of tumor radius evolution with free ($0\%$), $0.7\%$, and $1\%$ agarose concentrations as the fitting group, and the rest of the data as the comparison (or testing) group. We use the relative error between the simulated radii $R(t)$ and the experimental data $R_i$ as the metric of the goodness of fit, where the relative error is defined as $err=\sum_i (|R(t_i)-R_i|)/R_i$. Since we do not have {\it a priori} knowledge for the values of parameters, we first perform a coarse search over a logarithmic range of the parameters to identify approximate parameter ranges and then employ a finer grid-search to minimize $err$. The best-fit parameters with free, $0.7\%$, and $1\%$ gel concentration data for both compressible and incompressible models are listed in Table \ref{table1} in SM. \ref{nondimparam}.  

Using the tumor-associated parameters (i.e., all the parameters except $c_H$), we are able to fit the model to the tumor radii data for other gel concentrations  by only changing the parameter $c_H$ ($c_{H,0.3\%}$, $c_{H,0.5\%}$, $c_{H,0.8\%}$, $c_{H,0.9\%}$), and the optimal fit curves are shown in Figure \ref{FittingRadius}A. 
The bands shows the supremum and infimum envelope of the tumor radii by varying the parameter values within $10\%$ of the best fit. The simulations are in excellent agreement with the experimental data, which indicates that tumor growth is suppressed by the external gel confinement and the equilibrium tumor size is smaller when the agarose concentration is higher.

In the absence of external confinement (free growth case), our model predicts exponential growth at early stages, following by a gradual decrease in tumor growth speed that tends to become linear at later growth stages, consistent with proliferation being confined to a thin rim of cells at the spheroid boundary. This result differs qualitatively from previous models following Greenspan \cite{greenspan1976growth} where
unconfined tumor spheroids could still stop growing and reach an equilibrium size when the proliferation close to the tumor boundary perfectly compensates death close to the tumor center. The difference comes from our particular new feedback function that does not allow an arbitrary apoptosis rate \cite{greenspan1976growth}. We will discuss this further in Sec. \ref{model_analysis}.

To further demonstrate the effectiveness of our model, we employ the best-fit parameters in Table \ref{table1} in SM. \ref{nondimparam} obtained from the gel confinement experiment and predict the tumor radii evolution in the gel release experiment in \cite{helmlinger1997solid} where the gel is removed at a certain time. The only parameter to be determined by fitting with the gel release experimental data is the gel release time (when we set $c_H=0$ in simulations). With the obtained gel release times $T = 32$ and $T = 26$ for $0.7\%$ and $1\%$ gel concentrations, respectively, our model accurately predicts the tumor radii evolution at both growth stages before and after gel removal, as shown in Fig.~\ref{FittingRadius}B. Given that all experimental data points remain within $10\%$ of the simulation curve bands even with the potential discrepancies in cell line properties between two sets of experiments (gel confinement and gel release), it exhibits compelling evidence for the effectiveness of our model.

In the best-fit parameter set (Table \ref{table2} in SM. \ref{nondimparam}), we find the rearrangement rate $\beta=0$ to be optimal. To make sure our parameter inference is robust given the non-convex $err$ landscape with multiple local minima, we investigate the distribution of parameter sets including parameter combinations that gives relative errors $err$ no greater than $10\%$ of the best-fit error. The distribution for each parameter is summarized in the violin plots (Supplemental Figure \ref{ViolinHelm}). One can see that the median of $\beta$ is still close to $0$, suggesting very slow rearrangement activities for the mouse LS174T spheroid in \cite{helmlinger1997solid}. 

We further fit our model using data from \cite{Montel2021}, where human HT29 spheroid are grown subject to an applied pressure instead of spatial confinement. In particular, an applied pressure was induced through an osmotic shock at the surface of the spheroids and controlled by the osmotic stress in the surrounding medium with added dextran (Fig.~1 Bottom in \cite{Montel2021}). We extend our model to simulate the tumor growth under these conditions  by applying the external pressure boundary condition Eq.\eqref{eq:bc}, where $F_{ext} := P_{ext}$. We identify the model parameters as shown in Table \ref{table2} in SM. \ref{nondimparam} using the same methodology as in the gel confinement case, and simulate the growing tumor in a surrounding medium with osmotic pressure ranging from $0$ Pa, $500$ Pa, $2000$ Pa, to $5000$ Pa, as shown in Figure \ref{FittingRadius}C. Similar to the gel confinement case, there is excellent agreement with the experimental results that indicate that tumor growth is restrained by the applied external pressure. The broader shaded bands obtained by varying the parameter values within $10\%$ indicates the high sensitivity of tumor radius to the model parameters in the external pressure case. See Table \ref{table2} in SM. \ref{nondimparam} for the best-fit parameter set and the violin plots in Supplemental Figure \ref{ViolinPrl} for their robustness. It is noteworthy that in this case the inferred rearrangement rate $\beta$ is non-zero ($\beta = 0.06$ for the incompressible case and $\beta = 0.08$ for the compressible case) in contrast to the case with a confining gel.

Both the compressible and incompressible models fit the experimental data well. As shown in Table \ref{table3} in the Supplemental Material \ref{AICc scores}, while the relative errors are slightly smaller for the compressible model, the Aikake information criterion corrected for small sample sizes (AICc) predicts that the incompressible model is more likely to explain the data because the compressible model contains the additional parameter $K$. 

\subsubsection{Analysis of model predictions for the two experiments}
\label{model_analysis}
Now we look closely at the best fit parameters obtained from the two experiments of tumor growth under gel confinement (in Table \ref{table1}) and external pressure (in Table \ref{table2}). The most significant difference between two groups of best fit parameters is the value of the tissue rearrangement rate $\beta$, which is predicted (by both compressible and incompressible models) to be zero in the gel confinement experiment and to be positive ($\beta>0$) in the external pressure experiment with different cancer cell lines. The difference in parameter values might be attributed to the potential different cell line properties in two experiments, or more interestingly, to the different response of the tumor spheroids to the different external environments. To determine  what causes the differences in  
$\beta$ would need more systematic experimental investigation.

The predicted values of the elastic bulk modulus $K$ are also somewhat different  for gel confinement ($K=10$) and external pressure ($K=30$) experiments, but both suggesting a relatively small bulk modulus for the compressible tissue (compared to $K\rightarrow \infty$ for incompressible case). Similarly, the other parameters (such as $\eta$, $k$, $\gamma_c$ and $L$) differ somewhat between the two types of experiments, but are not 
substantially distinct. 

\begin{figure}[b]
\centering
  \scalebox{0.55}[0.55]{\includegraphics{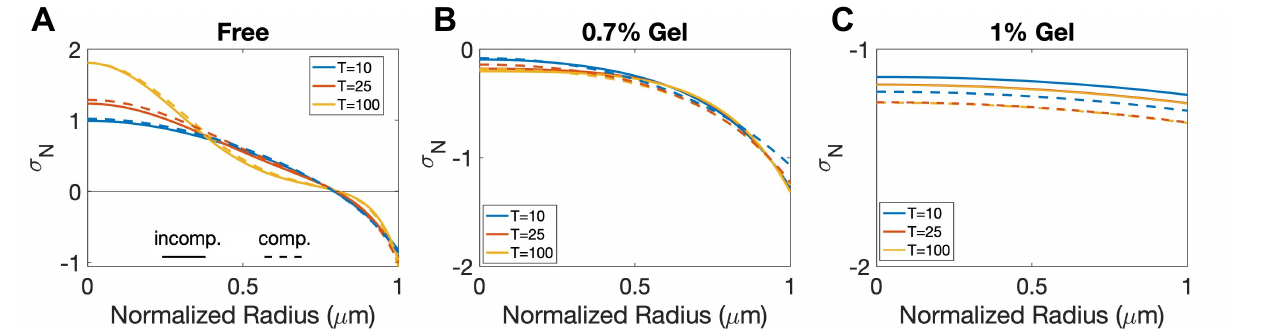}}
    \caption{The normal stress average $\sigma_N$ distributions at different times for A) free case ($F_{ext}=0$), B) $0.7\%$ gel concentration, and C) $1\%$ gel concentration. The solid and dashed lines show the results of the incompressible and compressible models, respectively.} 
    \label{FittingGelStressInvariant}
\end{figure}

Then, we ask if the new model generates stress distributions that are similar to those observed in previous work. As we found previously \cite{yan2021stress}, without an external gel (free growth), the normal stress average $\sigma_N$ is more tensile in the core region of tumor, but more compressive near the tumor boundary (Fig.~\ref{FittingGelStressInvariant}A). When the external gel is present, $\sigma_N$ becomes compressive within the entire tumor (Fig.~\ref{FittingGelStressInvariant}B and C), similar to that found in \cite{yan2021stress}. These results are also qualitatively consistent with the experimental observation of brain tumors in mice \cite{nia2016solid} where tensile or compressive stresses are distributed within the tumor depending on the surrounding environment and compression near the tumor interface. As the gel concentration increases, the tumor stresses become more compressive, but also more uniform along the radial direction, as shown in Fig.~\ref{FittingGelStressInvariant}C. The normal stress averages for the case with external pressure have very similar behaviors as in the gel confinement case (see Supplemental Figure \ref{ParameterStudyBeta}C, for example).

\begin{figure}[h]\centering
  \scalebox{0.55}[0.55]{\includegraphics{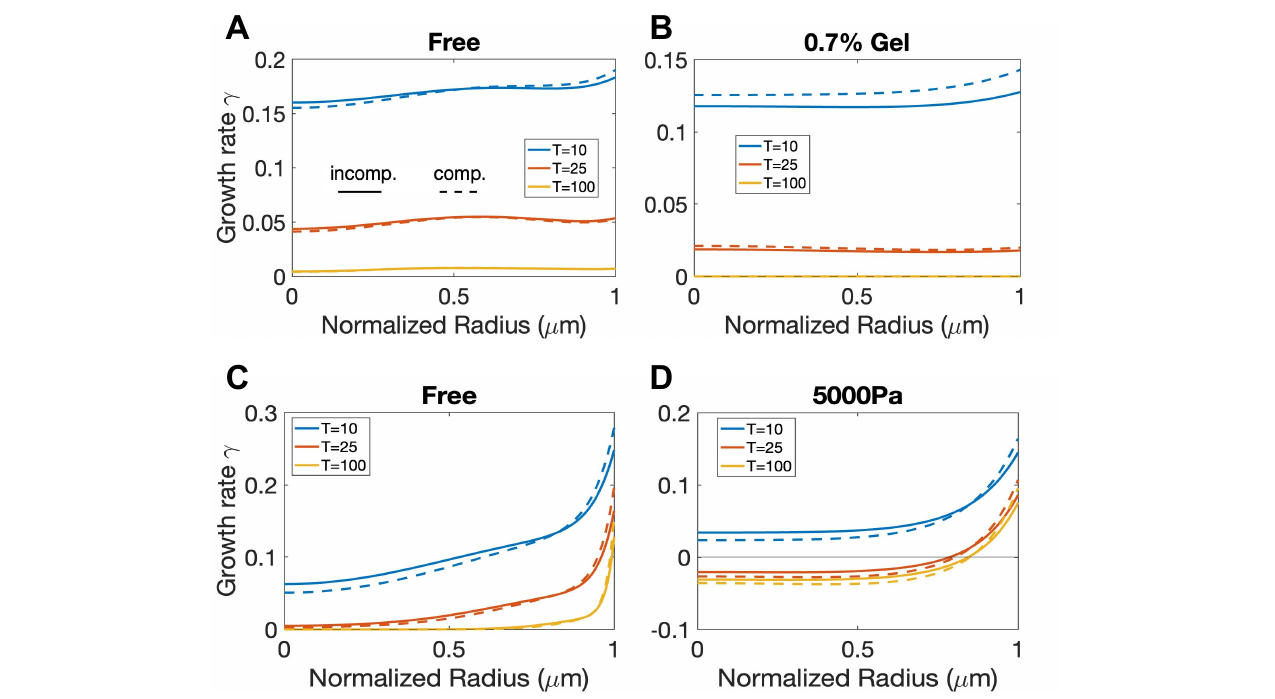}}
    \caption{The volumetric growth rate $\gamma$ distributions at different times for A) free case ($F_{ext}=0$), B) $0.7\%$ gel concentration, and C) free case in Parameter Table \ref{table2}, and D)  $5000$ Pa. The solid and dashed lines correspond to results using the incompressible and compressible models, respectively. } 
  \label{FittingGrowthRate}
\end{figure} 

In contrast to the stress distributions, the distributions of volumetric growth rates $\gamma$ can be very different for the two types of experiments. For \cite{helmlinger1997solid} with the best-fit rearrangement rate $\beta=0$, the volumetric growth rate is quite uniform inside the tumor region and will decrease overall as the tumor grows. Figs.~\ref{FittingGrowthRate}A and B predict the growth rate distributions without and with external confinement, respectively. For the free-growing case, our new model predicts that the spheroid will continue to grow with positive but decreasing growth rate (Fig.~\ref{FittingGrowthRate}A), which is different from our previous work \cite{yan2021stress} and other Greenspan-type models \cite{araujo2004linear,Wei2023elasticmodel} where an equilibrium size is guaranteed due to a volume-sink zone emerges at the center of the tumor. With external confinement, the distribution of growth rates will decrease and become uniformly zero (Fig.~\ref{FittingGrowthRate}B). This is essentially different from previous works \cite{araujo2004linear,yan2021stress,Wei2023elasticmodel} where isochoric rearrangement (or equivalently directed division) is necessary to stop stress from diverging over time even though the tumor size reaches equilibrium. The divergence of the elastic stress for the previous models results from a non-zero velocity field at steady state tumor sizes where the proliferating cells close to the boundary move inward to compensate the loss of volume due to cell death. Interestingly, we have not found any experimental observation to validate this non-vanishing inward velocity field. Strikingly, in our new model, since there is no growth anywhere in the tumor interior at equilibrium, stress dissipation is no longer necessary. 

With positive rearrangement rates $\beta>0$ (Figs.~\ref{FittingGrowthRate}C and D), the growth rate $\gamma$ becomes less uniform. In particular $\gamma$ is significantly larger near the tumor boundary but smaller near the tumor spheroid center (Fig.~\ref{FittingGrowthRate}C). Moreover, in the presence of the applied external pressure, $\gamma$ can become negative in the core region of the tumor and only remains positive near the tumor boundary (Fig.~\ref{FittingGrowthRate}D), even when the tumor reaches its equilibrium size. The result is similar to the Greenspan scenario with an inward tissue flow due to positive growth rate close to the boundary but negative growth rate close to the center. We emphasize this substantial difference in volumetric growth rates between two experiments should be attributed to the different values of tissue rearrangement rate $\beta$ rather than the external stimuli, which can be confirmed by the more detailed parameter study for $\beta$ in the next section (see Fig.~\ref{ResultsBeta}). We also note that the predicted growth rates of spheroid growth in the free and externally applied pressure cases is consistent with the distributions of cell proliferation and apoptosis in Fig.~2 from \cite{Montel2021}.

\begin{figure}\centering
  \scalebox{0.55}[0.55]{\includegraphics{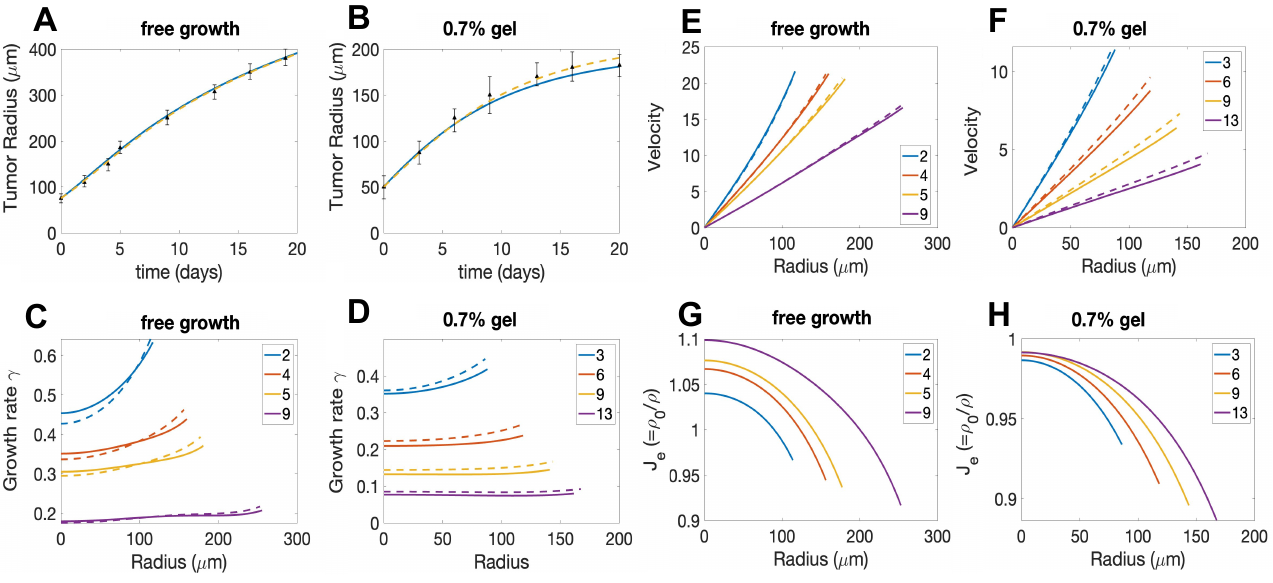}}
    \caption{The radii (A,B), the growth rates (C,D), the velocities (E,F) and elastic volume variation (G,H) for the models (solid-incompressible, dashed--compressible) using the best fit parameters.
     A) and B) show the radius evolution from T = 0 to T = 20 for free growth and 0.7\% gel confinement, respectively, with error bars from the experimental data. C) and D) depict the growth rates $\gamma$ in the tumor spheroid at different times for free growth (C) and 0.7\% gel confinement (D). E) and F) show the cell velocities $v$ in the spheroid for the free growth (E) 0.7\% gel confinement (F); G) and H) present the corresponding elastic volumetric variations $J_e$. } 
  \label{DiffComAndIncom}
\end{figure}

Since both compressible and incompressible models fit the experimental data well, we investigate the parameter values predicted by two models for the two experiments. The predicted parameter values by compressible and incompressible models are almost the same for the external pressure experiment (despite the difference in bulk modulus $K$); while the predicted values for the gel confinement experiment show noticeable disparity in the nutrient uptake rate $\gamma_c$ and diffusion coefficient $D$ (see also the violin plots in Supplement Figure \ref{ViolinHelm}). To gain insight on how these changes in parameters, and the bulk modulus $K$, are compensating one another, we investigate the evolution of tumor radius, growth rate distribution, velocity and elastic volumetric variation for the gel confinement experiment in Fig.~\ref{DiffComAndIncom}. Although the normal stress averages for compressible and incompressible cases are very similar (Fig.~\ref{FittingGelStressInvariant}A and B), the elastic volumetric variation $J_e$ is non-uniform in the compressible case and has about a 10\% smaller value at the spheroid boundary than in the spheroid center due to the relatively small $K$ (recall that $J_e=1$ in the incompressible case). In the compressible case, the increase in nutrient uptake $\gamma_c$ is accompanied by an increase in $D$,
 which makes nutrient penetrate farther into the tumor spheroid. This, combined with density, pressure and elastic energy variations, results in somewhat more heterogeneous volumetric growth rates $\gamma$ than those for incompressible case (Figure \ref{DiffComAndIncom} C and D). However, because of the nonuniform densities in the compressible case, these different growth rates ($\gamma$) result in similar cell velocities between the compressible and incompressible cases (Figure \ref{DiffComAndIncom} E and F), which is approximately the (weighted) integration of $\gamma$ from the tumor center to the boundary. Consequently, both compressible and incompressible models can successfully fit the experimental data for tumor radius evolution (Fig.~\ref{DiffComAndIncom}A and B). To get a better handle on how the parameters influence growth through biochemomechanical feedback, we perform a more extensive parameter study in the next section.

\subsection{Parameter Study}
\subsubsection{Effect of Mechanical Feedback}
In section \ref{subsec:nondim}, we have introduced the specific forms for the chemical energy density $\mathcal{E}_c=\frac{k}{2}c^2$ and the rescaling factor $\eta(c,\boldsymbol{\sigma})=\eta c$ in the chemomechanical feedback on growth \eqref{eq:gamma_feedback}, which yield the non-dimensional growth rate 
\begin{equation}
\gamma = \Big(\frac{1}{2}k \rho c^2 + \big(K(J_e-1)-J_e^{-1}W\big)\Big)\eta c,
\end{equation}
where $k$ is the chemical energy coefficient, which measures the strength of the chemical energy relative to the mechanical energy. Large $k$ implies weak mechanical feedback effect on the growth rate, whereas small $k$ implies a strong mechanical feedback effect. 

\begin{figure}
\centering
  \scalebox{0.55}[0.55]{\includegraphics{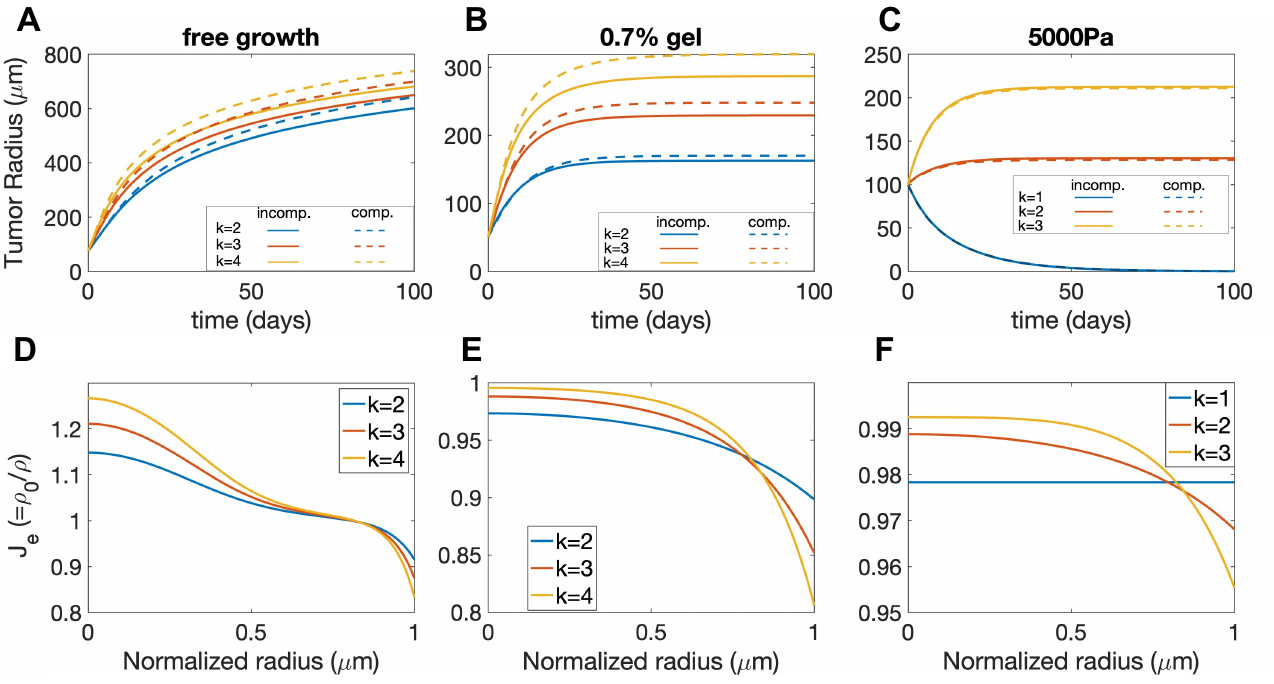}}
    \caption{Study of the parameter $k$, which represents the chemical energy coefficient that influences the proliferation rate. Figures A), B), and C) show the evolution of tumor along the time (days) for the free case, with gel confinement ($0.7\%$ gel), and with an applied static external pressure ($5000$ Pa). The solid and dashed lines indicate results using the incompressible and compressible models, respectively. Figures D), E), and F) show the elastic volumetric variations at time $T = 100$ for the compressible model.} 
  \label{Resultsk}
\end{figure}

We now investigate the effect of $k$ on the tumor growth in Fig.~\ref{Resultsk}. As expected, tumor growth slows down and the equilibrium tumor size is smaller (for the gel confinement and external pressure cases) when $k$ decreases due to the stronger mechanical feedback effect. More interestingly, when an external pressure is applied, the tumor starts shrinking and eventually the radius tends to zero when $k$ decreases below a certain level ($k=1$ in Fig.~\ref{Resultsk}C).  Tumor shrinkage is not observed for the gel confinement case. This result suggests that stronger mechanical feedback effect may induce tumor shrinkage in the presence of externally applied stress. This will be discussed further below. 
In addition, as $k$ decreases (e.g., stronger mechanical feedback), the elastic volumetric variation $J_e$ becomes more uniform (Fig.~\ref{Resultsk} D-F), the stress is more uniformly distributed
(Supplementary Fig.~\ref{ParameterStudyk} A-C), and the growth rate is smaller (Supplementary Fig.~\ref{ParameterStudyk} D-F).

\subsubsection{Elastic Relaxation in Tumor Dynamics}
\begin{figure}\centering
\scalebox{0.55}[0.55]{
\includegraphics{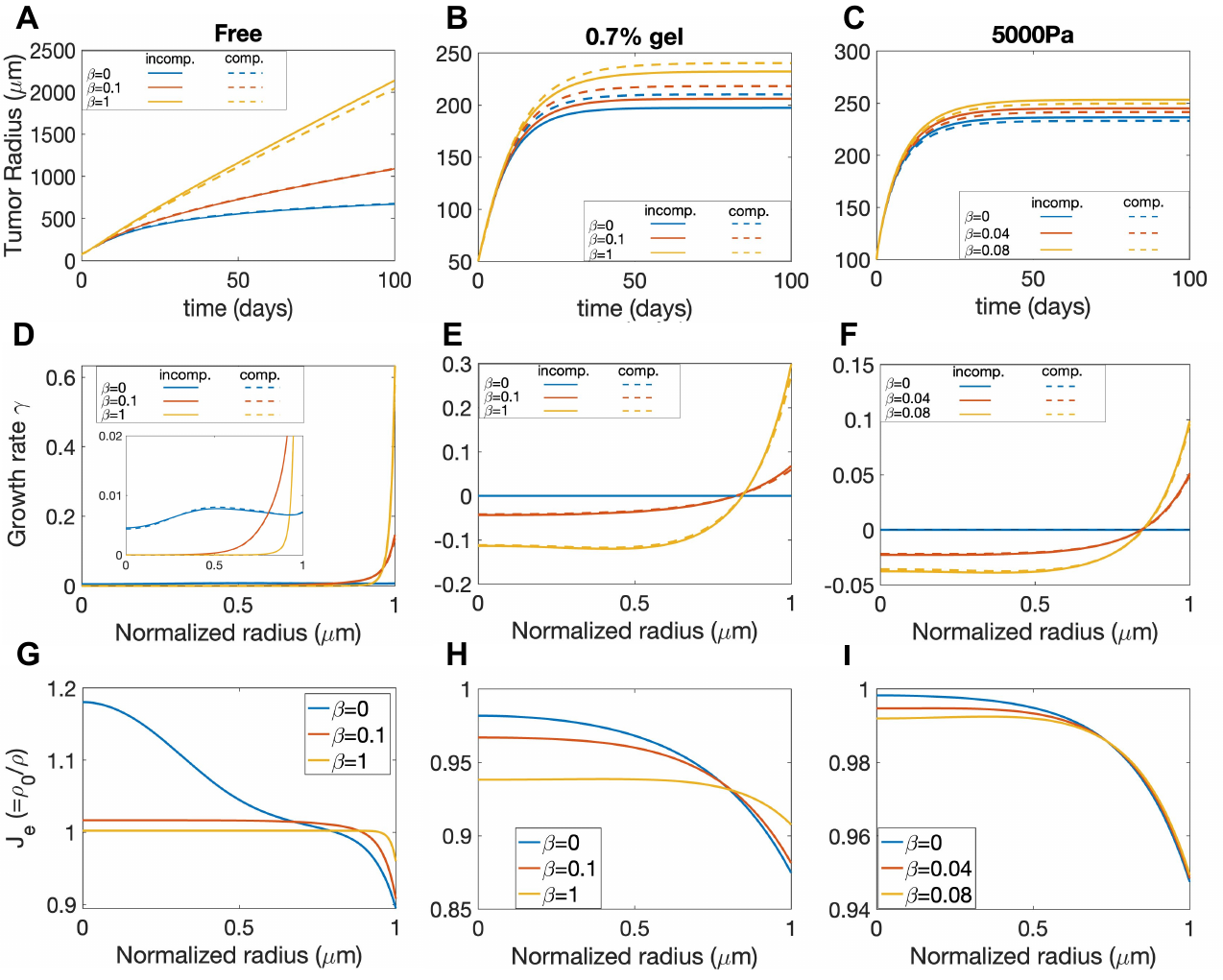}}
 \caption{Study of the relaxation parameter $\beta$ for the free case, gel confinement (0.7\% gel), and an applied external pressure (5000 Pa). A),B), and C) show tumor evolution over time (days) for each case. D), E), and F) depict the volumetric growth rate distributions in the normalized radial direction at T = 100. The solid and dashed lines correspond to the incompressible and compressible model results, respectively. G), H), and I) display elastic volumetric variation at T = 100 for the compressible model. 
  }
\label{ResultsBeta}
\end{figure}
Cancer cells can undergo significant rearrangement of surrounding tissues to facilitate tumor growth, invasion, microenvironment modulation, therapeutic resistance, and metastasis, which underscores its significance in cancer progression and treatment outcomes \cite{cox2011remodeling,winkler2020concepts,yuan2023extracellular}. The rearrangement can effectively relax the accumulated elastic stress \cite{legoff2016mechanical}, which is characterized by the tissue rearrangement rate $\beta$ that dissipates the elastic energy in our model. When fitting the experimental data of tumor size progression, we have found that $\beta$ can be zero (for the gel confinement experiment) or positive (for the external pressure experiment), and they can provide distinct growth patterns (Fig.~\ref{FittingGrowthRate}B and D). However, the distinct growth rate profiles can be also influenced by other model parameters, and other tumor growth behaviors can be also influenced by $\beta$. Therefore, we now investigate the sole effect of $\beta$ on tumor growth with free boundary, gel confinement and external pressure by only varying $\beta$ values.

First of all, since the growth rate is down-regulated by the elastic stress via mechanical feedback and tissue rearrangement can effectively relax the stress (Fig.~\ref{ParameterStudyBeta}), a larger tissue rearrangement rate $\beta$ promotes the tumor growth. Consequently, in the free growth case (agarose concentration is zero), the dynamics of the tumor radius evolution undergoes a transition from a logarithmic-like growth mode to a linear growth mode as the rearrangement rate $\beta$ increases (Fig.~\ref{ResultsBeta} A), consistent with proliferation becoming confined to an increasingly narrow region near the spheroid boundary. While the tumor grows infinitely large in the free growth case, an equilibrium tumor size exists in both the gel confinement and external pressure cases (Figs.~\ref{ResultsBeta} B and C). Moreover, the equilibrium tumor size increases with larger rearrangement rate $\beta$. 

Secondly, the distribution of volumetric growth rate $\gamma$ for the case with $\beta=0$ can be qualitatively different from the case with $\beta>0$. In the free growth case (Fig.~\ref{ResultsBeta} D), when $\beta=0$, $\gamma$ at $T=100$ is very small and is distributed more uniformly along the radial direction. When $\beta>0$, $\gamma$ is almost zero for most of the tumor but becomes very large near the tumor boundary, suggesting that the growth concentrates near the boundary for large $\beta$, consistent with the transition in growth mode observed in Fig.~\ref{ResultsBeta}. In the gel confinement and external pressure (Fig.~\ref{ResultsBeta} F) cases where an equilibrium tumor size exists (Fig.~\ref{ResultsBeta} E and F), $\gamma$ at equilibrium is uniformly zero within the entire tumor when $\beta=0$, while $\gamma$ is negative in the core region and positive near the boundary when $\beta>0$. The results demonstrate that the equilibrium growth rate $\gamma$ can change sign when $\beta>0$ and there is a source of external compression. 

Furthermore, through the modulation of stress, $\beta$ also affects the elastic volumetric variation $J_e$ (in the compressible case). In the absence of external confinement or pressure, the tumor core region is significantly stretched ($J_e>1$) while the boundary region is compressed ($J_e<1$) for negligible $\beta$ ($\beta=0$); as $\beta$ increases, the elastic stretch or compression is effectively relaxed and $J_e$ becomes more uniformly close to 1, as shown in Fig.~\ref{ResultsBeta} G. In the presence of a confining gel (Fig.~\ref{ResultsBeta} H) or external pressure (Fig.~\ref{ResultsBeta} I), the tumor is overall compressed ($J_e<1$) due to the external compression; the tumor boundary region is more compressed than the core region, and this nonuniformity is alleviated with increasing $\beta$. 


\subsubsection{External Forces on Tumor Equilibrium Size}
To be able to predict the size of tumor is important since tumor size correlates positively with critical prognostic factors and negatively affects survival rates of patients \cite{saha2015tumor}. 
We have observed in the previous sections that the tumor can undergo infinite growth, grow to a steady state or even shrink due to mechanical feedback and external mechanical stress. To determine the key factors that control the distinct behaviors for tumor growth, we now focus on the growth of an incompressible tumor ($J_e=1$) in the presence of a confining gel and external pressure. Moreover, we only consider the equilibrium tumor size in the case of $\beta=0$ since we found that for $\beta=0$ the volumetric growth rate $\gamma$ will eventually become uniform zero within the entire tumor when the tumor reaches its equilibrium size. 
Based on this observation, we obtain the following nonlinear.  nonlocal system for the equilibrium tumor radius $R$ and the elastic stress (or equivalently the elastic deformation strains $fe_\theta$ and $fe_r={fe_\theta}^{-2}$)
\begin{align}
    P &= \sigma_{rr}+ \int_R^r \frac{2}{s}(\sigma_{rr}-\sigma_{\theta\theta})ds +F_{ext}\label{EReq1_main}, \\
    \gamma &= 0 \implies W+P-\frac{I_1}{d} = \frac{k}{2}c(R,r)^2, \label{EReq2_main}
\end{align}
where Eq.~\eqref{EReq1_main} comes from the force balance equation and Eq.~\eqref{EReq2_main} is the condition for zero growth rate. Given the equilibrium radius $R$, an analytical expression of the equilibrium solutions $c(R,r)$ to the reaction diffusion equation \eqref{eq:c_evol} can be easily obtained. Combining the boundary conditions of $fe_\theta$ and $c(R,r)$, we can numerically compute $R$ and $fe_\theta(r)$ by a shooting method based on Eqs. \eqref{EReq1_main} and \eqref{EReq2_main}. From the numerical results, we find that $fe_\theta(r)\in (0,1]$ for all radii $r$ with different $k$, $c_H$ and $P_{ext}$ values, as shown in Fig.~\ref{ResultsEqR} C and D for instance. Further details about the shooting method and analysis can be found in Supplemental \ref{EQRsec}. 

More interestingly, we can obtain a result on the equilibrium radius $R$ by simply analyzing \eqref{EReq1_main} and \eqref{EReq2_main} at the boundary. Given the boundary condition $c(R,R)=1$, we can obtain the following equation at $r=R$:
\begin{equation}
\frac{k}{2}-\big(W+\sigma_{rr}-\frac{I_1}{d}\big)=F_{ext}.
\end{equation}
Since $W\geq0$ and $\sigma_{rr}-I_1/d=\frac{2}{3}(1/fe_\theta^4-fe_\theta^2)\geq 0$ given $fe_\theta \in (0,1]$, we can conclude that $F_{ext}\leq\frac{k}{2}$ for a valid solution of the above force balance equation at $r=R$ to exist. When the applied external pressure is too large, namely $P_{ext}>\frac{k}{2}$, it is impossible for the internal stress to balance with the external pressure and hence the tumor radius will shrink to zero under a constant pressure $P_{ext}$. In contrast, when there is a confining gel, even if the external traction $F_{ext}>\frac{k}{2}$ at some point, which leads to a decrease in $R$, the tumor will not continue to shrink since $F_{ext}$ also decreases with $R$ following Eq. \eqref{chforceEQ}. Consequently, force balance will eventually hold for some $R>0$. 

On the other hand, considering Eqs. \eqref{EReq1_main} and \eqref{EReq2_main} at $r=0$ yields
\begin{equation}
\frac{k}{2}c(R,0)^2 -\int_R^0 \frac{2}{s}(\frac{1}{fe_\theta^4}-fe_\theta^2)ds = F_{ext}. 
\end{equation}
Since the concentration $c(R,0)>0$, the coefficient $k>0$ and $fe_\theta \in (0,1]$, the left hand side of the equation is positive, which can be regarded as the expected external traction for force balance provided that a valid solution with $0<R<\infty$ exists. However, when $F_{ext}\equiv 0$ ($c_H=0$ or $P_{ext}=0$), force balance cannot hold and a finite equilibrium radius $R$ does not exist. Furthermore, since $F_{ext}=0$ cannot cancel with the positive effective internal stress, the tumor radius $R$ will grow to infinity, as we have seen from the previous results for free growth case. 

\begin{figure}\centering
  \scalebox{0.55}[0.55]{\includegraphics{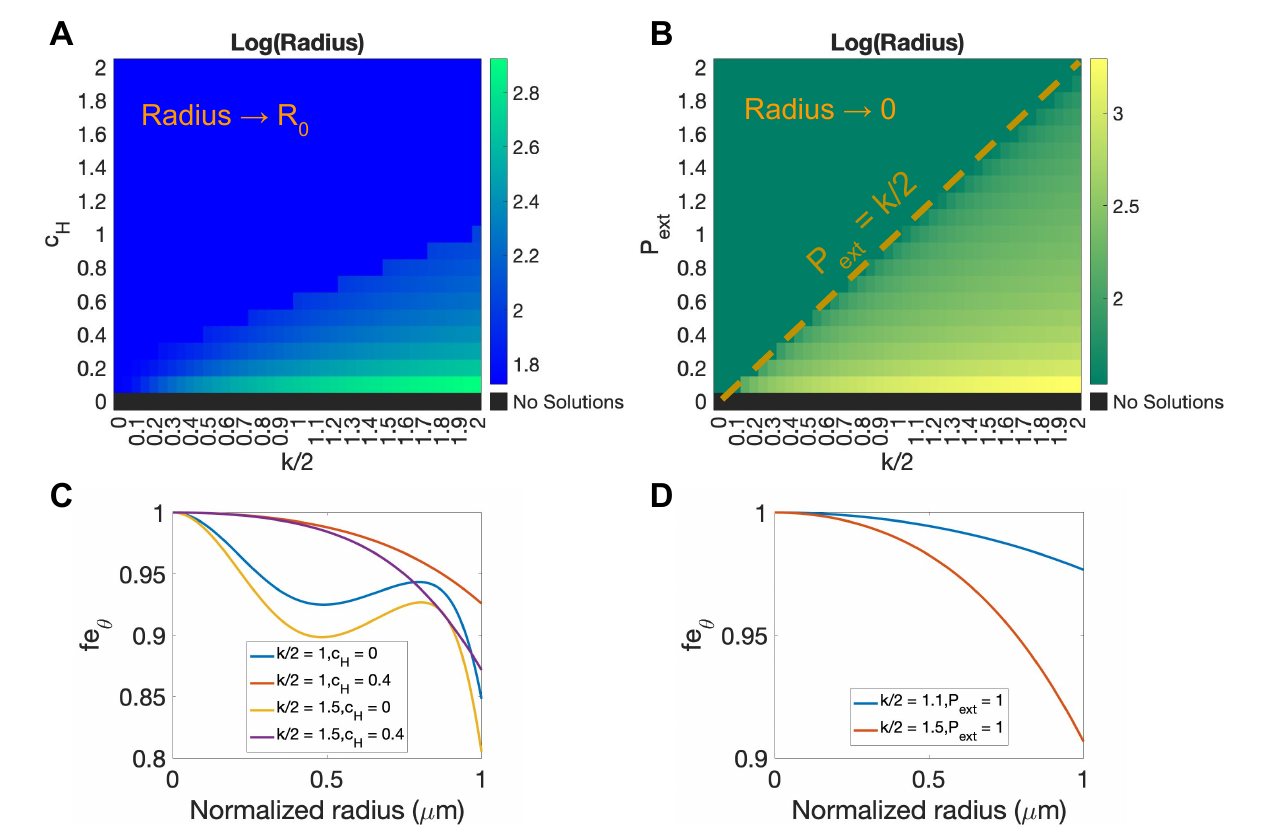}}
    \caption{Numerically predicted equilibrium radii (log scale) for the incompressible model without elastic relaxation $\beta=0$ under A) gel confinement ($c_H$) and B) external pressure ($P_{\rm ext}$) with $L=65, \gamma_c=0.7$ as functions of chemical energy coefficient $k/2$. C) and D) show numerical results of $fe_\theta$ (0 to 1) under varying $k, c_H, P_{ext}$.} 
  \label{ResultsEqR}
\end{figure}

Next, we numerically determine equilibrium tumor radii for varying intensities of mechanical feedback $1/k$ and strengths of the external mechanical stimuli ($C_H$ for the gel confinement case and $P_{ext}$ for the external pressure case) using the shooting method. The results are shown in Figs.~\ref{ResultsEqR}A and B. In the absence of a confining gel or external pressure ($C_H=0$ or $P_{ext}=0$), the tumor grows indefinitely and an equilibrium size does not exist. When $C_H>0$ or $P_{ext}>0$, the tumors grow to an equilibrium size and the equilibrium size decreases as the the intensity of mechanical feedback increases (as $k$ decreases) and the strength of external stimuli increases (as $C_H$ or $P_{ext}$ increases). In the gel confinement case, tumor growth can be significantly suppressed and the equilibrium tumor size is very close to its initial size when $C_H$ is large enough and $k$ is small enough. However, the equilibrium tumor radius does not decay and is always above its initial size. 
In contrast, in the external pressure case, the tumor size can  shrink below its initial size and eventually decay to zero if $P_{ext}$ is large enough and $k$ is small enough. As indicated above (and clearly seen in Fig.~\ref{ResultsEqR}B), the critical threshold for the transition of the equilibrium size from nonzero to zero is given by 
$P_{ext} = \frac{k}{2}$.

\subsubsection{Effect of Tissue Compressibility and other parameters}
\begin{figure}
\centering
  \scalebox{0.55}[0.55]{\includegraphics{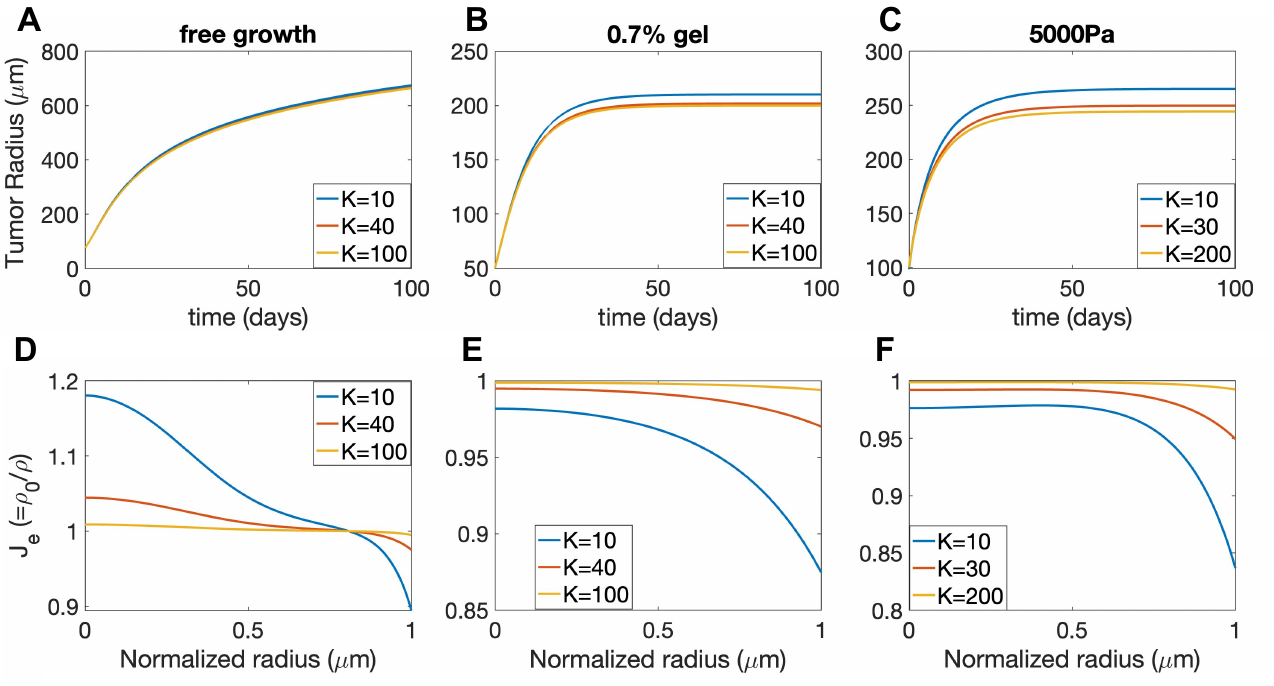}}
    \caption{Study of the parameter $K$, which represents the bulk modulus and influences tissue compressibility. Figures A) $\&$ B) and C) show the evolution of tumor along time (days) for the free case, with gel confinement ($0.7\%$ gel), and with an applied static external pressure ($5000$ Pa). Figure D)  The solid lines indicate incompressible results and dashed lines indicate compressible results. E) and F) show the elastic volumetric variations at time $T = 100$ for the compressible model.} 
  \label{ResultsK0}
\end{figure}

Cell compressibility is a biomechanical marker for assessing cancer's malignant transformation and metastatic potential \cite{fu2021measurement}, and it is represented by the bulk modulus $K$ in our model. As $K$ increases, the tumor becomes less compressible, requiring larger external forces or pressures to induce volumetric deformations. Conversely, the tumor becomes more compressible as $K$ decreases, allowing it to undergo larger volumetric deformations under the same external conditions. To investigate the effect of $K$, we perform multiple simulations of tumor growth varying $K$ values while keeping other parameters as the best fit values in Tables \ref{table1} and \ref{table2}. 
The tumor equilibrium radii are not very sensitive to the tissue compressibility $K$ (Fig.~\ref{ResultsK0} A-C), although smaller $K$ values result in larger tumor equilibrium sizes. The modulation of $R$ by $K$ might be the effect of nonlinear advection flow as discussed in \cite{Wei2023elasticmodel}, rather than the effect of the mechanical feedback given that different $K$ values do not notably change the normal stress average and the equilibrium volumetric growth rate (Fig.~\ref{ParameterStudyK0}). 
However, the compressibility $K$ can significantly alter the elastic volumetric variation $J_e$, or equivalently the reciprocal of the local cell density $\rho_0/\rho$, as shown in Figure \ref{ResultsK0} D-F. As expected, when $K$ is smaller, $K$ is more nonuniform along the radial direction with the spheroids being most compressed at the spheroid boundary. As $K$ increases, $J_e$ becomes more uniform and closer to 1, approaching the incompressible tumor case. 

 The other parameters involved in the reaction-diffusion process of nutrients such as the uptake rate $\eta_c$ and the diffusion coefficient $D$ can also modulate tumor growth. As illustrated in Figs.~\ref{ParameterStudygammac} and \ref{ParameterStudyL}, as the uptake rate $\eta_c$ increases, the nutrient concentration decreases and tumor growth is inhibited. Conversely, as $D$ increases, the nutrient level increases and growth is promoted.

\section{Discussion}
In the present paper, we have proposed a framework for the chemomechanical regulation of growth based on thermodynamics of continua and growth-elasticity. In particular, by considering both the elastic and chemical energy of the system and using a variational approach, we derived a novel formulation for chemomechanical feedback that incorporates an energy-dissipating stress relaxation and biochemomechanical regulation of the volumetric growth rate. We further applied our model to study tumor growth in a confining gel  \cite{helmlinger1997solid} and an external pressure \cite{Montel2021}. By fitting with available experimental data, we demonstrate the effectiveness of our model in describing biochemomechanical regulation of tumor growth under different external mechanical environments. We also investigated the importance and effect of the model parameters (such as the tissue rearrangement rate, tissue compressibility, strength of mechanical feedback and external mechanical stimuli) on the growth of tumor spheroids through numerical simulations. 

In contrast to previous biomechanical models that either combine different or ad-hoc approaches for growth regulation into one model, our model provides a systematic view of biochemomechanical regulated growth, where the mass-volumetric growth and mass-conserving rearrangement are separately constrained and coupled with a reaction-diffusion process of the chemical field. We note the framework can be easily extended to account for regulation of processes from non-diffusive chemical species (e.g., through intracellular processes). Multiple chemical species and their corresponding energies can also be added into the current system. We have shown that the model is able to reproduce experimental data on growth rates of tumor spheroids and exhibits a wider range of growth-rate distributions than previous models. An essential difference from previous growth models is that the tumor continues to grow without reaching an equilibrium size when no external compression or confinement is considered. External confinement is required for the tumor to reach a steady size. We cannot conclude whether this new system is more effective to describe large-scale tissue growth than previous models as this requires more experimental data such as the kinematics of the entire growing domain.

On mass-conserving tissue rearrangement, we have previously developed both compressible \cite{Wei2023elasticmodel} and incompressible \cite{yan2021stress} models that combined growth-elasticity theory with an adaptive reference map. Both our previous and current models, when linearized, recover Maxwell viscoelastic fluid constitutive laws \cite{Ranft-2010-pnas,Streichan-2018-elife} (see Supplemental \ref{linearmodel} for details). Advantages of our new framework are that the relaxation has a clear geometric meaning and is straightforward to implement. The rearrangement here simply means there is a local irreversible change of tissue shape, but not size, given by the deviatoric part of the (Eulerian) growth-rate tensor. Previously, we needed to pair the evolution of growth tensor and the adaptive reference map to account for a mass-conserving rearrangement making the active geometric change of local tissue harder to interpret in non-spherical geometries.
Although this is a continuum framework, the deviatoric part of the (Eulerian) growth-rate tensor can be compared with the events of cell-cell T1 transitions in 2-dimensional tissues \cite{david2014tissue}. Previous results for the compressible tissues \cite{Wei2023elasticmodel} show that increasing the rearrangement relaxes the stress field, thus results in more uniform density field. The rearrangement in the current model also exhibits this effect. 

With both mass-conserving tissue rearrangement and biomechanical regulation on volumetric growth, the new model produces both similar and distinct results from the model in \cite{yan2021stress}. For the whole tissue size, increasing the rearrangement relaxes the stress overall thus increases the tissue expansion, which is shown in both the current and previous models \cite{yan2021stress}. Compared to the previous model, parameters such as strength of mechanical feedback and external mechanical stimuli all have similar effects on the growth and mechanics on the growing tissue. A new feature of the model is that the growth rate pattern changes according to the rearrangement rate. When the rearrangement rate is relatively large, the growth distribution is similar to previous works where there is more growth close to the boundary. With smaller rearrangement, this trend is decreased and the growth rate is more uniform. Whether tissue rearrangement has such an impact on growth patterning needs further experimental validation. For example, one may check cell proliferation and apoptosis patterns when adhesion molecule endocytosis along cell-cell junctions is perturbed. These molecules are partially responsible for tissue-level rearrangements \cite{katsuno2020endocytosis}. 

Lastly, we comment on the numerical solutions to the current framework. As this is an Eulerian model, we solve the velocity field together with the elastic deformation field, instead of solving the coordinate maps and growth tensor field in a Lagrangian frame \cite{budday2015size,garcia2017contraction}. As the velocity field is not explicitly involved in the mechanical equilibrium equation, we take the material time derivative of the mechanical equilibrium equation to reveal the velocity and we discretize the spatial derivatives using second order finite differences. This is similar to the idea in \cite{Jones2000} to solve velocity from a linear elastic system by taking the time derivative of the original stress balance equation that involves displacement field. Although the resulting equation is nonlinear in the elastic deformations, it is linear in the velocity field.
This enabled us to design a simplified numerical scheme where we first solve a linear system for velocity, and use the velocity field to evolve the elastic fields. This approach can be extended to multiple dimensions, by adding extra features to track/capture the boundary and resolve the bulk-boundary coupling. This will be done in the future work.

\section*{Acknowledgments}
The authors acknowledge partial funding from the National Science Foundation- Division of Mathematical Sciences (NSF-DMS) under grants DMS-1953410 (JL), DMS-2309800 (JL), DMS-1763272/Simons Foundation (QN594598, JL), DMS-2012330 (MW), and DMS-2144372 (MW). CW acknowledges partial funding from National Natural Science Foundation of China under grants 12371392 and 12431015. NO acknowledges the Development and Promotion of Science and Technology Talent Project (DPST) in Thailand. JL additionally acknowledges partial funding from the National Institutes of Health (NIH) grant P30CA062203 for the Chao Family Comprehensive Cancer Center at UC Irvine. MW acknowledge partial funding from the National Institute of General Medical Sciences (NIGMS) of NIH under award number R01GM157590.

\begingroup
\small
\vskip2pc
\bibliographystyle{unsrt}
\bibliography{ref.bib}
\endgroup

\vfill
\eject

\appendix

\noindent \textcolor{black}{{}\textbf{\Large{Supplemental Materials for 
Chemomechanical regulation of growing tissues from a thermodynamically-consistent framework and its application to tumor spheroid growth}}}
\section{Non-dimensionalization and optimal parameter values}
\label{nondimparam}

We consider the following non-dimensionalization:
\begin{align*}
& \tilde{\mathbf{x}}=\frac{\mathbf{x}}{l}, \quad \tilde{t}=\frac{t}{\tau}, \quad \tilde{\gamma}_c=\gamma_c \tau, \quad \tilde{\gamma}=\gamma \tau, \quad \tilde{\beta}=\beta \tau, \quad \tilde{\alpha} = \alpha\frac{l^2}{\mu \tau},~ \tilde{\rho}=\frac{\rho}{\rho_0}, \quad \tilde{c}=\frac{c}{c_0} \\
& \quad \tilde{D} = D\frac{\tau}{l^2}, \quad \tilde{k}=k\frac{\mu}{\rho_0 c_0^2}, \quad \tilde{\eta} = \eta(\mu \tau c_0),~ \tilde{K}=\frac{K}{\mu}, \quad \tilde{c}_H = \frac{c_H}{\mu}, 
 \quad \tilde{W} = \frac{W}{\mu},
\end{align*}
where $l$ is a characteristic length, $\tau$ is a characteristic time, and the tilde quantities are dimensionless. Here we take $l = 1$ $\mu m$ and $\tau = 1$ day. The non-dimensionalized system becomes equivalent to the original equations in the main text when we remove all tildes and set $\mu = 1$, and $c_0 = 1$.

\begin{table}[ht]
\begin{center}
\begin{tabular}{ |l|c|c| } 
 \hline
 \centering \bf{Parameters} & \bf{Incomp. Values}&\bf{Comp. Values}\\
 \hline
 $R_0 $ : tumor initial radius ($\mu m$) & \multicolumn{2}{|c|}{Depends on the experimental 
 data} \\ 
  \hline
  $K$ : bulk modulus & - & 10\\
 \hline 
 $\beta $ : elastic relaxation & 0 & 0  \\
 \hline
 $\eta$ : rescaling factor of & 0.8 & 0.7\\the volumetric growth rate $\gamma$ &  &  \\
 \hline
  $k$ : chemical energy constant & 2.5 & 2.5\\
 \hline
  $\gamma_c$ : positive constant uptake rate & 0.8 & 1.3\\
 \hline
 $D$ : diffusional coefficient of nutrient  & $65^2$ & $90^2$ \\ 
 \hline
 $c_{H,0.7\%}$ : shear modulus of $0.7\%$ gel & 0.32 & 0.34\\
 \hline
 $c_{H,1\%}$ : shear modulus of $1\%$ gel & 0.82 & 0.87\\
 \hline
\end{tabular}
\end{center}
\caption{Values of model parameters in Figure 2A.}
\label{table1}
\end{table}

\begin{table}[ht]
\begin{center}
\begin{tabular}{ |l|c|c| } 
 \hline
 \centering \bf{Parameters} & \bf{Incomp. Values}&\bf{Comp. Values}\\
 \hline
 $R_0 $ : tumor initial radius ($\mu m$) & \multicolumn{2}{|c|}{Depends on the experimental 
 data} \\ 
  \hline
  $K$ : bulk modulus & - & 30\\
 \hline 
 $\beta $ : elastic relaxation & 0.06 & 0.08  \\
 \hline
 $\eta$ : rescaling factor of & 1.2 & 1.2\\the volumetric growth rate $\gamma$ &  &  \\
 \hline
  $k$ : chemical energy constant & 3.5 & 3.5\\
 \hline
  $\gamma_c$ : positive constant uptake rate & 1.2 & 1.2\\
 \hline
 $D$ : diffusional coefficient of nutrient  & $70^2$ &  $70^2$ \\ 
 \hline
$P_{ext, 500\text{ Pa}}$ : compression pressure for $500$ Pa & 0.35 & 0.35 \\
  \hline
$P_{ext, 2K\text{ Pa}}$ : compression pressure for $2000$ Pa& 0.45 & 0.45  \\
  \hline
$P_{ext, 5K\text{ Pa}}$ : compression pressure for $5000$ Pa & 0.65 & 0.65\\
\hline
\end{tabular}
\end{center}
\caption{ Values of model parameters in Figure 2C.}
\label{table2}
\end{table}

\section{AICc scores}
\label{AICc scores}
\begin{table}[!h]
\begin{center}
\begin{tabular}{ |c|c|c|c|c| } 
\hline
  & \multicolumn{2}{|c|}{Gel Confinement} & \multicolumn{2}{|c|}{Pressure}\\
   & AICc & ERR & AICc & ERR \\
 \hline
 Incompressible & \bf{8.9098} & 0.8318 & \bf{4.1657} & 0.4977 \\
 \hline
 Compressible & 11.1107 & 0.7738 & 5.1822 & 0.4276\\
\hline
\end{tabular}
\end{center}
\caption{AICc scores comparing between incompressible and compressible models}
\label{table3}
\end{table}
The corrected Akaike information criterion (AICc) \cite{bumham2002model} is used to determine which model best fits the data. The model with the lower AICc score explains the data better. The AICc score is calculated as
\begin{equation}
    \text{AICc} = n \text{Log}(\frac{\text{ERR}}{n}+\frac{2mn}{n-m-1})
\end{equation}
where $n$ is the number of observed data points, ERR is the relative error, and $m$ is the number of estimable parameters in the model. The number of data points using for Figure\ref{FittingRadius}A and Figure\ref{FittingRadius}C is $n = 28$ and $n = 44$, respectively. The AICc predicts that the incompressible model provides a better fit to the data, even though the overall error is smaller for the compressible model. This is because the compressible model contains an additional parameter (the bulk modulus $K$).

\section{Frame Invariance of the Dynamic System}\label{FrameInvariance}
 We show that the dynamic system of the deformation tensors is frame invariant even with tissue rearrangement. Consider an observer in a frame $\mathbf{x}^+=\boldsymbol\chi^+(\mathbf{X},t+)$ that connects to the frame $(\mathbf{x},t)$ via the transformation $\mathbf{x}^+(\mathbf{X},t+)=\mathbf{Q}(t)\mathbf{x}(\mathbf{X},t)+\mathbf{c}(t)$, where $\mathbf{Q}(t)$ is an arbitrary orthogonal rotational tensor ($\mathbf{Q}^{\text{T}}=\mathbf{Q}^{-1}$) and $t^+=t+\alpha$. Without loss of generality, we will assume $\alpha=0$ and $t^+=t$. 
Based on the transformation, we have $\mathbf{F}^+=\mathbf{Q}\mathbf{F}$. For the decomposition $\mathbf{F}=\mathbf{F}_e\mathbf{F}_g$, as the translation and rotation only act in the spatial coordinate system, we have $\mathbf{F}_g^+=\mathbf{F}_g$ but $\mathbf{F}_e^+=\mathbf{Q}\mathbf{F}_e$. 
Given that $(\nabla\mathbf{v})^+=\mathbf{Q}\nabla\mathbf{v}\mathbf{Q}^{\text{T}}+(d\mathbf{Q}/dt) \mathbf{Q}^{\text{T}}$, we can show that the dynamics of the geometric deformation gradient Eq.~(\ref{eq:F_evol}) is frame invariant:
\begin{equation}
    \frac{d \mathbf{F}^+}{dt} - (\nabla \mathbf{v})^+\mathbf{F}^+= \mathbf{Q}\Big(\frac{d \mathbf{F}}{dt} - (\nabla \mathbf{v})\mathbf{F}\Big)= \mathbf{0}.
\end{equation}
Since $J^+ = \det(\mathbf{F}^+)=\det(\mathbf{Q})\det(\mathbf{F})=J$ and $(\nabla\cdot\mathbf{v})^+=\tr(\mathbf{Q}\nabla\mathbf{v}\mathbf{Q}^{\text{T}})+\tr((d\mathbf{Q}/dt) \mathbf{Q}^{\text{T}})=\tr(\nabla \mathbf{v})=\nabla\cdot\mathbf{v}$ noticing that $(d\mathbf{Q}/dt) \mathbf{Q}^{\text{T}}$ is skew-symmetric, we can show the frame invariance of the dynamics of $J$ Eq.~(\ref{eq:J_evol})
\begin{equation}
\frac{dJ^+}{dt} = J^+(\nabla\cdot\mathbf{v})^+.
\end{equation}
Considering the Eulerian growth rate tensor $\boldsymbol \Gamma=\frac{\gamma}{d}\mathbf{I}+\boldsymbol{\Gamma}_D$ and $\tilde{\boldsymbol \Gamma} = \mathbf{F}_e^{-1}\boldsymbol \Gamma\mathbf{F}_e$, we have $\boldsymbol \Gamma^+=\mathbf{Q}\boldsymbol \Gamma \mathbf{Q}^{\text{T}}$ and the unaffected $\tilde{\boldsymbol \Gamma}^+=\tilde{\boldsymbol \Gamma}$. Then we have the frame invariant dynamics of $\mathbf{F}_g^+$
\begin{equation}
\frac{d \mathbf{F}_g^+}{dt} = \tilde{\boldsymbol\Gamma}^+\mathbf{F}_g^+,   
\end{equation}
and the frame invariant dynamics of $\mathbf{F}_e^+$
\begin{equation}
\frac{d \mathbf{F}_e^+}{dt} -(\nabla\mathbf{v})^+\mathbf{F}_e^+=\mathbf{Q}\Big(\frac{d \mathbf{F}_e}{dt} -\nabla\mathbf{v}\mathbf{F}_e\Big)=-\mathbf{Q}\boldsymbol\Gamma\mathbf{F}_e=-\boldsymbol\Gamma^+\mathbf{F}_e^+.
\end{equation}
Similarly with the dynamics of $J^+$, we can show that the dynamics of $\rho$, $J_e$ and $J_g$ is also frame invariant. Moreover, we can confirm that the evolution of the Finger deformation tensor Eq.~\eqref{eq:be_evol} is frame invariant:
\begin{equation}\label{eq:be_evol_invariant}
\frac{d\mathbf{B}_e^+}{dt} - (\nabla \mathbf{v})^+\mathbf{B}_e^+-\mathbf{B}_e^+(\nabla \mathbf{v}^\text{T})^+=\mathbf{Q}\stackrel{\triangledown}{\mathbf{B}}_e\mathbf{Q}^\text{T}=-\frac{2}{d}{\gamma}\mathbf{B}_e^+- \big(\boldsymbol{\Gamma}_D^+\mathbf{B}^+_e+\mathbf{B}^+_e\boldsymbol{\Gamma}^+_D\big),
\end{equation}
where $\mathbf{B}_e^+=\mathbf{Q}\mathbf{B}_e\mathbf{Q}^\text{T}$ and $\boldsymbol{\Gamma}_D^+=\mathbf{Q}\boldsymbol{\Gamma}_D\mathbf{Q}^\text{T}$ are the transformed finger deformation tensor and deviatoric Eulerian growth rate tensor respectively, and $\stackrel{\triangledown}{\mathbf{B}}_e=d\mathbf{B}_e/dt- \nabla \mathbf{v}\mathbf{B}_e-\mathbf{B}_e\nabla \mathbf{v}^\text{T}$ is the upper-convective time derivative. 

\section{Viscoelastic behavior of the linearized model}\label{linearmodel}
 We demonstrate the stress-relaxation behavior in the linearization of our nonlinear model. In the theory of small deformations, we assume that $\mathbf{F}=\mathbf{I}+\nabla\mathbf{u}$, $\mathbf{F}_g=\mathbf{I}+\mathbf{G}$ and $\mathbf{F}_e=\mathbf{I}+\mathbf{E}$, where the displacement $\mathbf{u}=\mathbf{x}-\mathbf{X}$, the growth increment tensor $\mathbf{G}$ and the elastic increment tensor $\mathbf{E}$ are small such that their norms are of order $\epsilon$ with $\epsilon\ll 1$. By considering the leading-order approximations, we obtain the linearized stress $\boldsymbol\sigma=\boldsymbol\sigma_D+\boldsymbol\sigma_p$:
\begin{align}
\boldsymbol\sigma_p &= K\tr(\mathbf{E})\mathbf{I}, \label{eq:pressure_linear} \\ 
\boldsymbol\sigma_D &= \mu ( \mathbf{E}+\mathbf{E}^{\text{T}}-(2/d)\tr(\mathbf{E})\mathbf{I}). \label{eq:stress_linear}
\end{align}

To derive the evolution of $\boldsymbol\sigma_D$ and  $\boldsymbol\sigma_p$, we first obtain the linearization of the elastic deformation tensor (\ref{eq:Fe_evol}) with $\boldsymbol{\Gamma}_D$ defined in \eqref{eq:gamma_D}:
\begin{equation}
\frac{\partial \mathbf{E}}{\partial t} = \nabla\mathbf{v}-\frac{\gamma}{d}\mathbf{I}-\beta\Big( \mathbf{E}+\mathbf{E}^{\text{T}}-\frac{2}{d}\tr(\mathbf{E})\mathbf{I}\Big),
\end{equation} 
where we have omitted the advection term $\mathbf{v}\cdot\nabla\mathbf{E}$ of smaller order. It yields the evolution of isotropic stress (or pressure)
\begin{equation}
\frac{\partial {\boldsymbol\sigma}_p}{\partial t}=K(\nabla\cdot{\mathbf{v}}-\gamma)\mathbf{I},
\end{equation}
and the stress relaxation behavior of viscoelastic materials for deviatoric stress
\begin{align}
\boldsymbol\sigma_D+\frac{1}{2\beta}\frac{\partial {\boldsymbol\sigma}_D}{\partial t}=\frac{\mu}{2\beta}\Big(\nabla \mathbf{v}+\nabla\mathbf{v}^{\text{T}}-\frac{2}{d}(\nabla\cdot\mathbf{v})\mathbf{I}\Big),
\end{align}
where $(\nabla \mathbf{v}+\nabla\mathbf{v}^{\text{T}})$ defines the strain rate, $\beta$ serves as half the rate of stress relaxation and the ratio $\mu/(2\beta)$ plays the role of "effective" viscosity. This Maxwell-like model of deviatoric stress matches exactly with the linearzation of our previous nonlinear elastic model with ``adaptive reference map" \cite{Wei2023elasticmodel}, where we describe the tissue rearrangement by the adaption of reference state to the current deformed state via $d\mathbf{X}/dt=\tilde{\beta}(\mathbf{x}-\mathbf{X})$ and the two relaxation coefficients are related by $\tilde{\beta}=2\beta$.

\section{Finding the Equilibrium Radius} \label{EQRsec}
For incompressible tumor, when $\beta = 0$, we found that $\gamma(r) \equiv 0$ in the entire tumor at equilibrium. Based on this observation, we obtain a system that connects the equilibrium radius $R$ and the solution of $fe_\theta$ as below
\begin{align}
& P-\frac{1}{fe_\theta^4} = \int_R^r \frac{2}{s}(\frac{1}{fe_\theta^4}-fe_\theta^2)ds +F_{ext},\label{EReq1}\\
& \frac{1}{2}(2fe_\theta^2+\frac{1}{fe_\theta^4}-3)+P = \frac{k}{2}c(R,r)^2+\frac{1}{3}(2fe_\theta^2+\frac{1}{fe_\theta^4}),\label{EReq2}
\end{align}
where the first equation comes from the force balance equation and the second equation comes from the fact $\gamma(r)\equiv 0$, and $c(R,r)$ is the analytical solution of the quasi-equilibrium reaction diffusion equation \eqref{eq:c_evol}, namely $L^2 \frac{1}{r^2}\frac{\partial}{\partial r}(r^2\frac{\partial c}{\partial r}) - \gamma_c c =0$, given by
\begin{equation}
    c(R,r) = \frac{R \sinh{(\frac{\sqrt{\gamma_c}r}{L})}}{r\sinh{(\frac{\sqrt{\gamma_c}R}{L})}},
    \label{eq:Cequi}
\end{equation}
where we use $L=65$ and $\gamma_c=0.7$ without loss of generality. The solutions of \eqref{EReq1} and \eqref{EReq2} satisfy the following boundary conditions: 
\begin{align*}
    fe_\theta &= 1 \text{ at } r=0,\\
    P-\sigma_{rr} &= F_{ext} \text{ at } r=R\\
    c &= 1 \text{ at } r = R.
\end{align*}
Letting $x$ be $fe_\theta^2$, we substitute $P$ from equation ($\ref{EReq2}$) to equation ($\ref{EReq1})$ as follows
\begin{equation}
    \frac{k}{2}c(R,r)^2 - \frac{1}{2}(2x+\frac{1}{x^2}-3)-\frac{2}{3}(\frac{1}{x^2}-x)-F_{ext} =  \int_R^r \frac{2}{s}(\frac{1}{x^2}-x)ds \label{EReq3}
\end{equation}
If we multiply equation (\ref{EReq3})  by $x^2$ and rearrange the terms, we obtain a cubic polynomial of $x$ as below
\begin{equation}
  H(x,r;R):=(-\frac{1}{3})x^3+\left(\frac{k}{2}c(R,r)^2+\frac{3}{2}-F_{ext}- \int_R^r \frac{2}{s}(\frac{1}{x^2}-x)ds\right)x^2 - (\frac{7}{6}) = 0, \label{EReq4}    
\end{equation}
where the coefficients depend on $r$ and $R$. In other words, given an equilibrium radius $R$, we can obtain the solution $x$ at position $r\in[0,R]$ based on the above equation. Based on our numerical calculations, we found that $fe_\theta$ monotonically decreases from the center of the tumor to the boundary when $\beta=0$. Furthermore, since we have $fe_\theta=1$ at $r=0$ and $fe_\theta>0$ for all $r$ based on its physical interpretation, we can conclude that $x \in (0,1]$ for $r\in[0,R]$. We also confirm this result $fe_\theta\in (0,1]$ by numerical simulations. 

Now let us consider the equation \eqref{EReq4} at $r=0$ 
\begin{equation}
    \frac{k}{2}c(R,0)^2 -\int_R^0 \frac{2}{s}(\frac{1}{x^2}-x)ds = F_{ext}. 
\end{equation}
If $F_{ext} = 0$, it is not possible to find an equilibrium solution for $0<R<\infty$ since the left hand side is positive given that $c(R,0)>0$ and $k>0$ and $x \in (0,1]$. Furthermore, the radius $R$ will increase indefinitely due to the positive external driving force expressed by the left hand side. 

Next, consider the equation \eqref{EReq3} at $r=R$, which yields
\begin{equation}
-\big[\frac{1}{2}(2x+\frac{1}{x^2}-3)+\frac{2}{3}(\frac{1}{x^2}-x)\big]=F_{ext}-\frac{k}{2}
\label{EqR:Pext_and_k}
\end{equation}
where we have used the fact $c(R,R)=1$ at $r=R$. Since $x \in (0, 1]$, the left hand side is non-positive. For a valid solution $x \in (0, 1]$ to exist, we would require that $F_{ext}\leq \frac{k}{2}$. If the constant external pressure is too large, i.e., $P_{ext}> \frac{k}{2}$, there does not exist an equilibrium radius $R$ and the radius $R$ will decrease to zero due to the larger external pressure. However, when gel confinement is present ($c_H>0$), an equilibrium radius will always exist because $F_{ext}$ depends on $R$ by the relation in Eq. \eqref{chforceEQ}. When $F_{ext}>\frac{k}{2}$, the radius $R$ decreases and meanwhile $F_{ext}$ also decreases, and the radius $R$ will reach an equilibrium once $F_{ext}$ is small enough such that the force balance in Eq. \eqref{EqR:Pext_and_k} holds for some $x\in(0,1]$.

Assuming an equilibrium radius $R$ exists, we use shooting method to find the appropriate value for $R$. Specifically, given an initial guess for equilibrium radius $R$, we can compute $x(r)$ from $r=R$ to $r=0$ and then update the radius $R$ by matching the boundary condition $x=1$ at $r=0$. In particular, we discretize along the radius by $r_i=i\Delta r$ ($i=0,1,\ldots,N$) with $\Delta r= R/N$ and use the right endpoint rule to approximate the integral. By considering Eq. \eqref{EReq4} with the integral from $r_j$ to $R$ for $j=N, \ldots, 0$, we can solve $x(r_j)$ sequentially. In other words, $x(r_j)$ depends on the values of {$x(r_i)$ for all $i=j+1,\ldots, N$ that have been obtained before. When solving the value of each $x(r_j)$, if no positive real root exists for Eq. \eqref{EReq4}, we will assume $x(r_j) = x(r_{j+1})$ and move on to compute $x(r_{j-1})$. While this assumption is not physical, it aids in obtaining numerical values of $x(r_j)$ for all $j$. If there are multiple real roots for Eq. \eqref{EReq4}, we will choose the one with the smallest positive value. Notice that, there could be critical numerical errors from approximating the integral when $r$ is small (close to zero) because of the singular factor $\frac{2}{s}$. Instead using the integrand directly, we can approximate $x \approx 1+\frac{1}{2}x_{rr}(r)r^2$ using Taylor's expansion around $r = 0$ using the condition $x=1$ at $r=0$. Then the integral is rewritten as 
$$\int_R^r \frac{2}{s}(-\frac{x_{rr}^3}{8}s^6-\frac{3x_{rr}^2}{4}s^4-\frac{3x_{rr}}{2}s^2) \frac{1}{(1+\frac{1}{2}x_{rr}s^2)^2} ds.$$
Since we have used the boundary condition $x(0)=1$, we eventually match the equation in Eq. \eqref{EReq4} rather than the boundary condition. In particular, in order to find the equilibrium tumor radius $R$, we perform the bisection method with the residual of the equation $H(R)$ in Eq. \eqref{EReq4}. When $H(R)<0$, it indicates that the given $R$ is smaller than the equilibrium. Otherwise, choose the other section for new interval until it converges. In our setup, we use this method when $r$ is less than $10^{-3}$ and the tolerance for the method is $10 ^{-16}$.

\section{Supplemental Figures}
\setcounter{figure}{0}
\renewcommand{\thefigure}{S\arabic{figure}}
\begin{figure}[!h]
\begin{center}
  \scalebox{0.5}[0.5]{\includegraphics{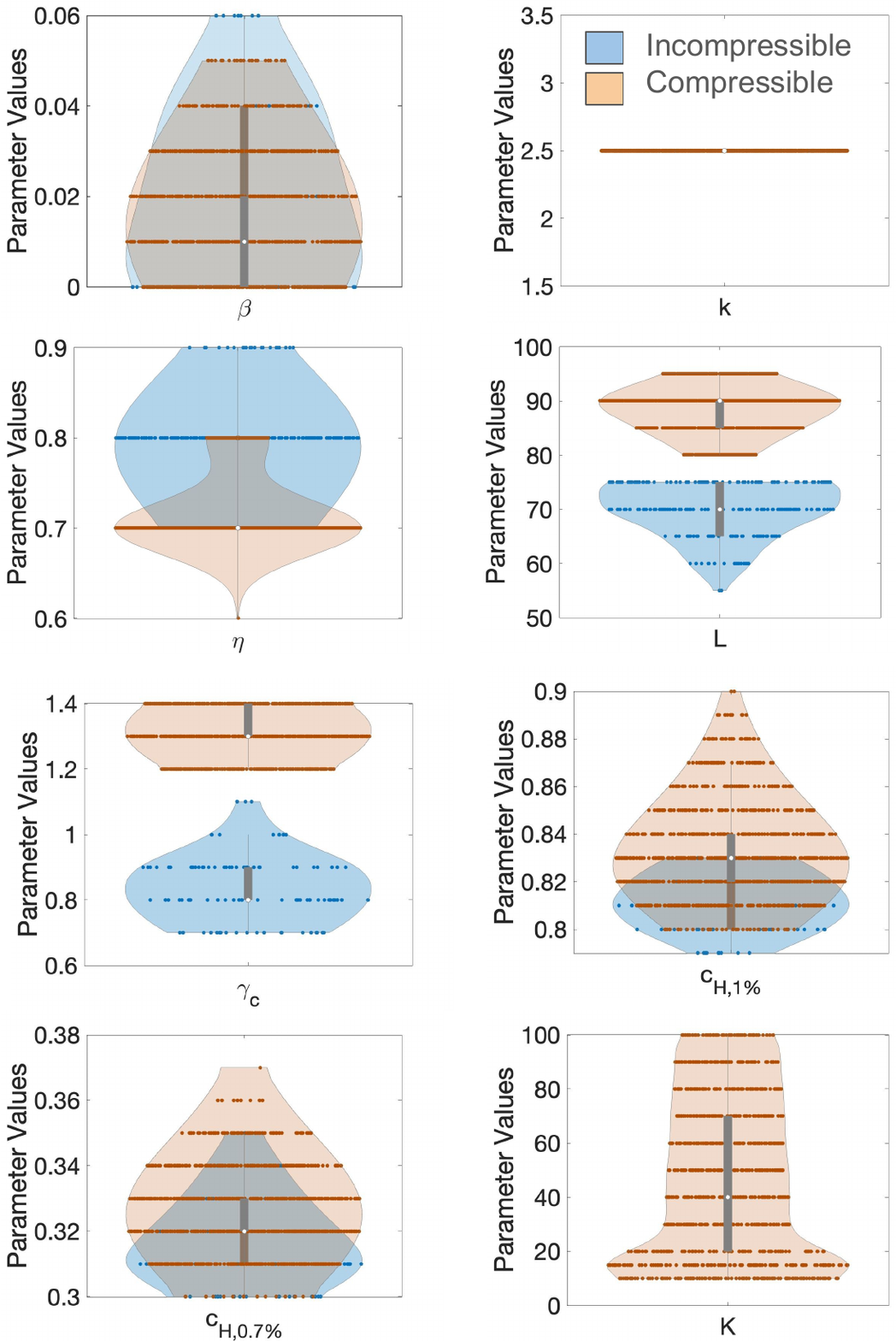}}
    \caption{Violin plots displaying the distribution of data sets that are within $10\%$ of the best fitting relative error from Figure \ref{FittingRadius}A. The plots represent each parameter where blue is for the incompressible model and red is for the compressible model. The width of the violin at any given point represents the density of data points at that value. Additionally, there is a central box-and-whisker plot, which provides summary statistics such as the median, quartiles, and outliers of the dataset within each category. Notably, the parameters take distinct values for the compressible and incompressible cases, particularly for $L=\sqrt{D}$ and $\gamma_c$. These disparities may rise from complex compensatory effects among other parameters, such as $K$ and $c_H$.} 
  \label{ViolinHelm}
  \end{center}
\end{figure}

\begin{figure}
\begin{center}
  \scalebox{0.5}[0.5]{

\includegraphics{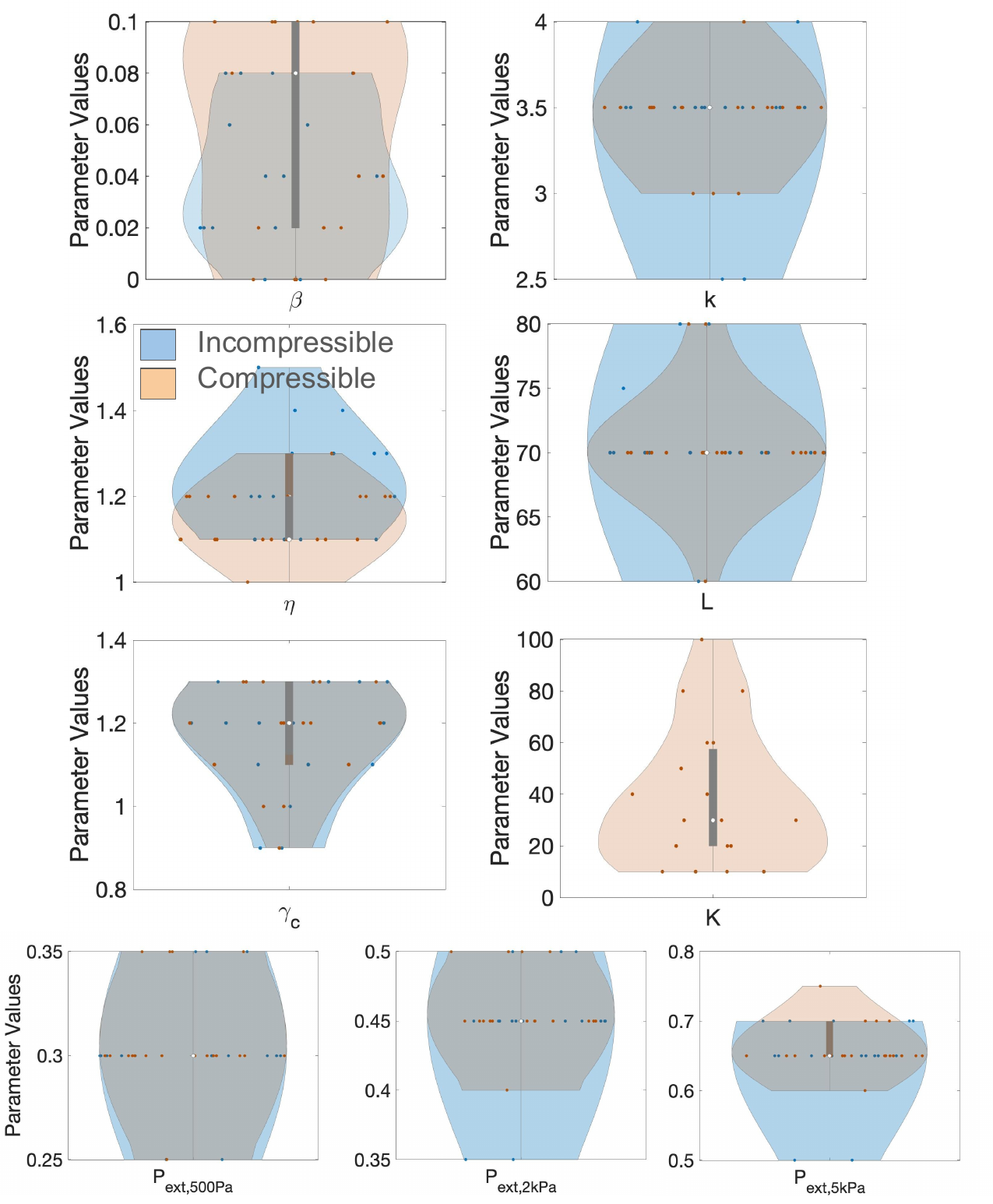}}
    \caption{Violin plots showing the distribution of data sets that are within $10\%$ of the best fitting relative error from Figure \ref{FittingRadius}C. The plots present each parameter where blue designates the incompressible model and red designates the compressible model. The width of the violin at any given point represents the density of data points at that value. Additionally, there is a central box-and-whisker plot, which provides summary statistics such as the median, quartiles, and outliers of the dataset within each category. These plots highlight the similarities of the best-fitting parameters between compressible and incompressible models.} 
  \label{ViolinPrl}
  \end{center}
\end{figure}

\begin{figure}[!h]\centering
  \scalebox{0.62}[0.62]{\includegraphics{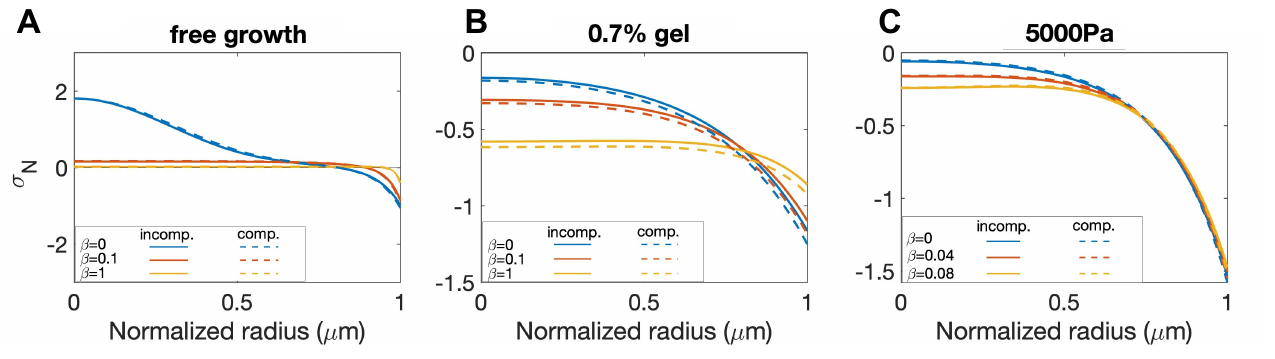}}
    \caption{Study of the parameter $\beta$ where A), B), and C) display the normal stress average $\sigma_N$ distribution at T=100 for free case, with gel confinement ($0.7\%$ gel), and with an applied static external pressure ($5000$ Pa), respectively.} 
  \label{ParameterStudyBeta}
\end{figure}
\begin{figure}[!h]\centering
  \scalebox{0.62}[0.62]{\includegraphics{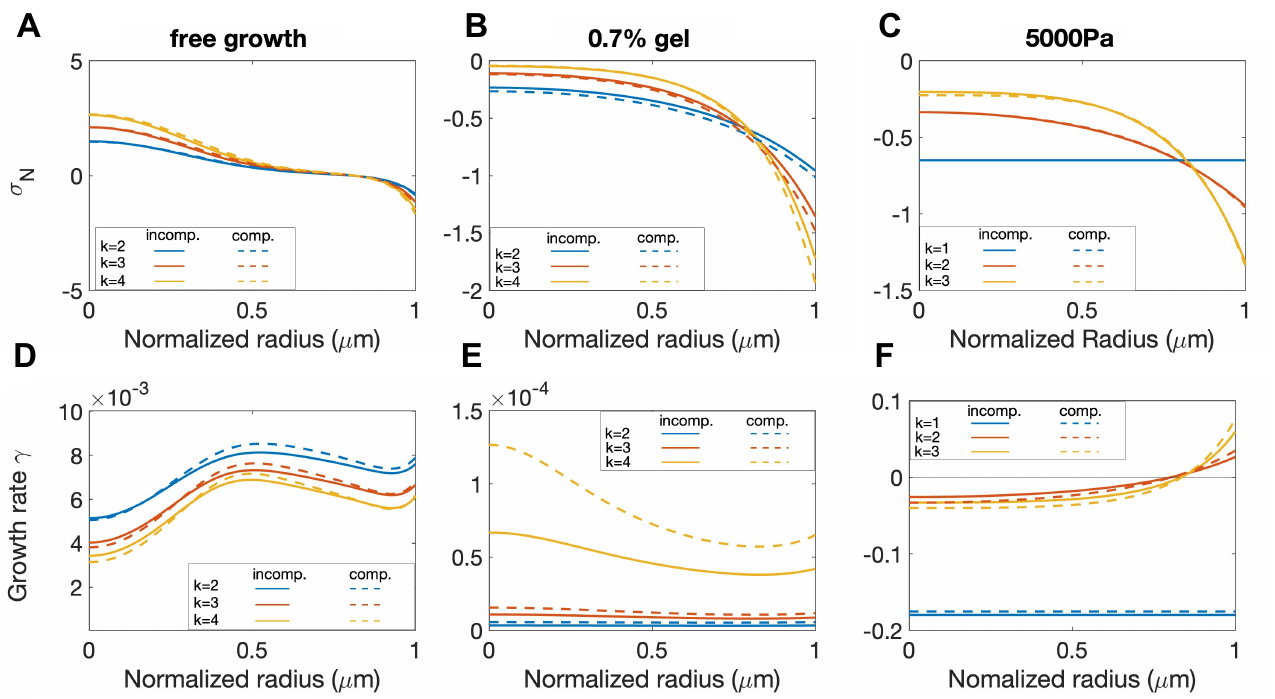}}
    \caption{Study of the parameter $k$ where A), B), and C) display the normal stress average $\sigma_N$ distribution at T=100 for free case, with gel confinement ($0.7\%$ gel), and with an applied static external pressure ($5000$ Pa), respectively. D), E), and F) display the growth rate $\gamma$ distribution at $T=100$ for free case, with gel confinement ($0.7\%$ gel), and with an applied static external pressure ($5000$ Pa), respectively. } 
  \label{ParameterStudyk}
\end{figure}
\begin{figure}[!h]\centering
  \scalebox{0.62}[0.62]{\includegraphics{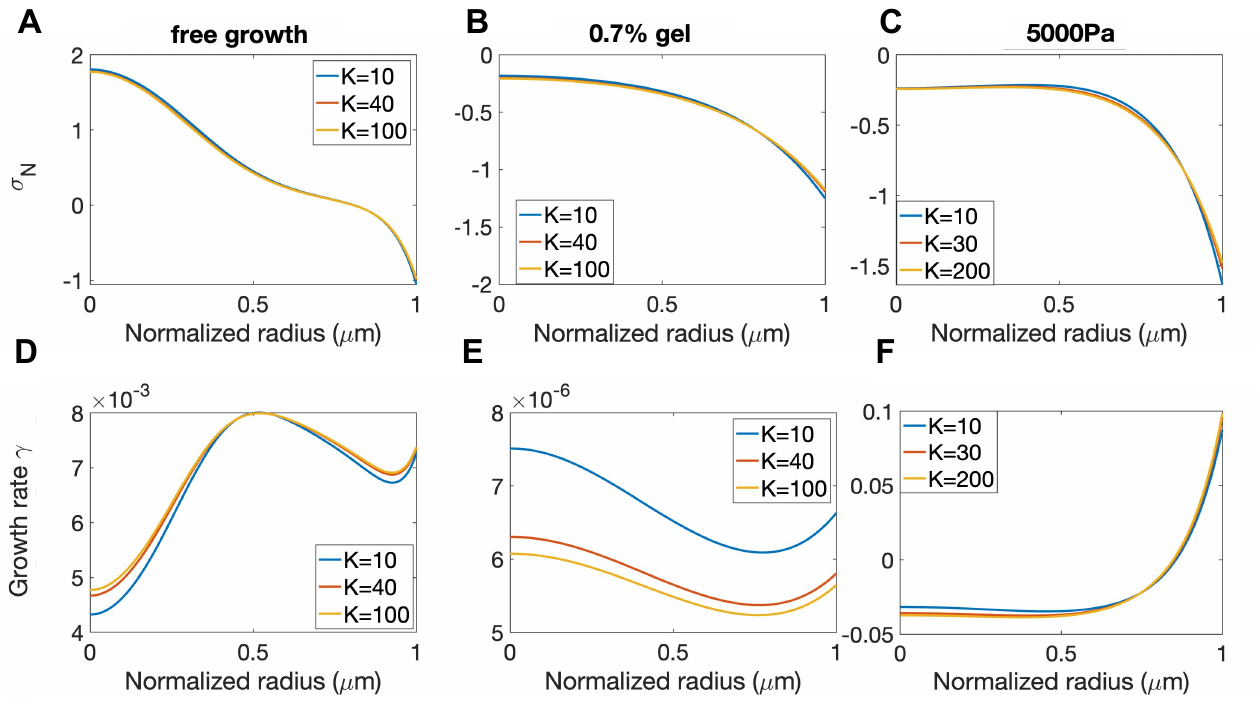}}
\caption{Study of the parameter $K$ where A), B), and C) display the normal stress average $\sigma_N$ distribution at T=100 for for free case, with gel confinement ($0.7\%$ gel), and with an applied static external pressure ($5000$ Pa), respectively. D), E), and F) display the growth rate $\gamma$ distribution at $T=100$for free case, with gel confinement ($0.7\%$ gel), and with an applied static external pressure ($5000$ Pa), respectively. }  \label{ParameterStudyK0}
\end{figure}

\begin{figure}[!h]\centering
  \scalebox{0.62}[0.62]{\includegraphics{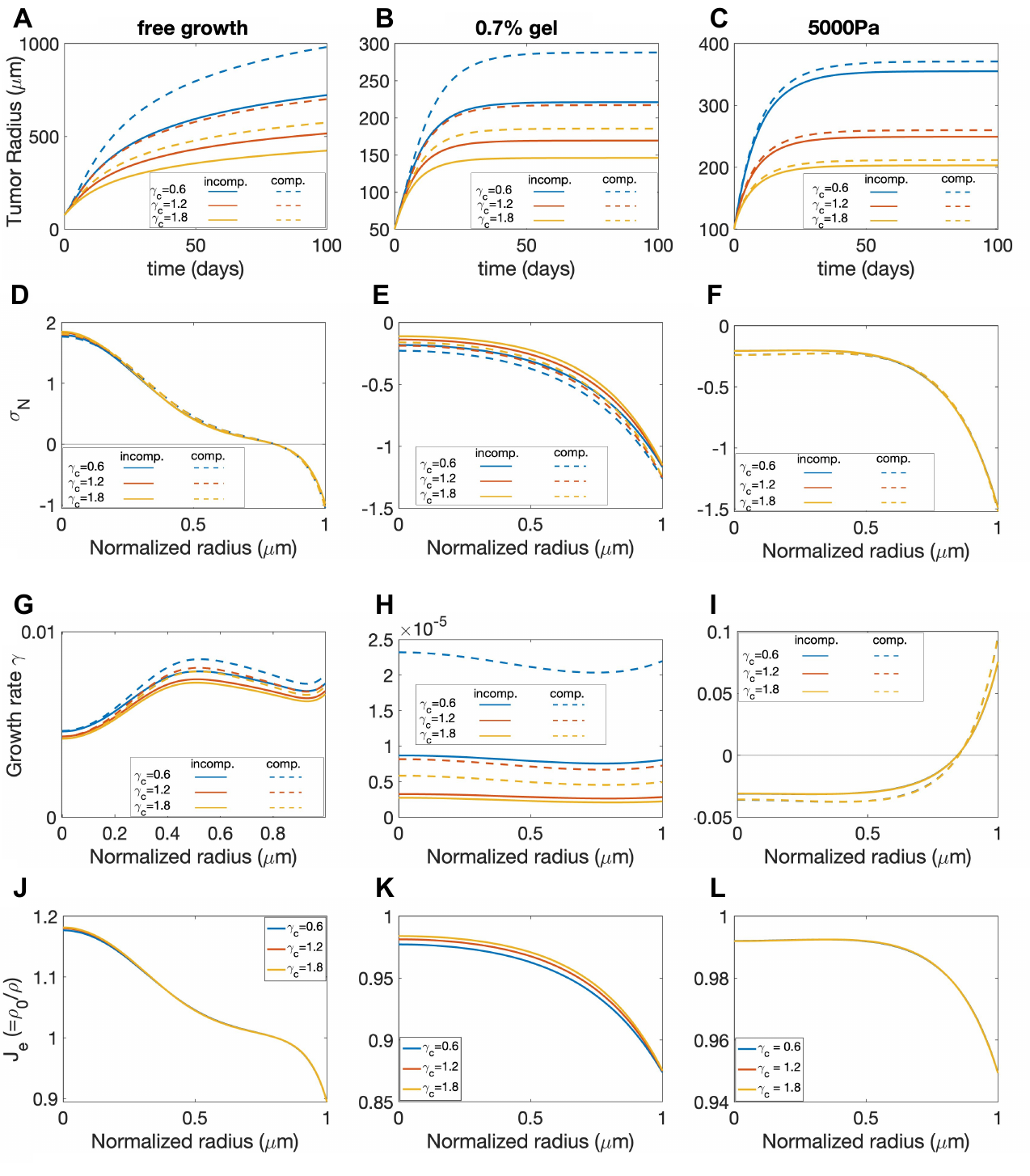}}
    \caption{Study of the parameter $\gamma_c$, which characterizes the uptake rate in our nutrient transportation equation for free case, with gel confinement ($0.7\%$ gel), and with an applied static external pressure ($5000$ Pa). Figures A) $\&$ B) and C) show the evolution of tumor for each case along time (days). Figures D) $\&$ E) and F) show the normal stress average $\sigma_N$ distributions in the normalized radial direction at time $T = 100$. Figures G) $\&$ H) and I) indicate the volumetric growth rate distributions in the normalized radial direction at time $T = 100$. The solid lines indicate incompressible results and dashed lines indicate compressible results. Figure J) $\&$ K) and L) display the elastic volumetric variation at time $T = 100$ for compressible model.} 
  \label{ParameterStudygammac}
\end{figure}
\begin{figure}\centering
  \scalebox{0.62}[0.62]{\includegraphics{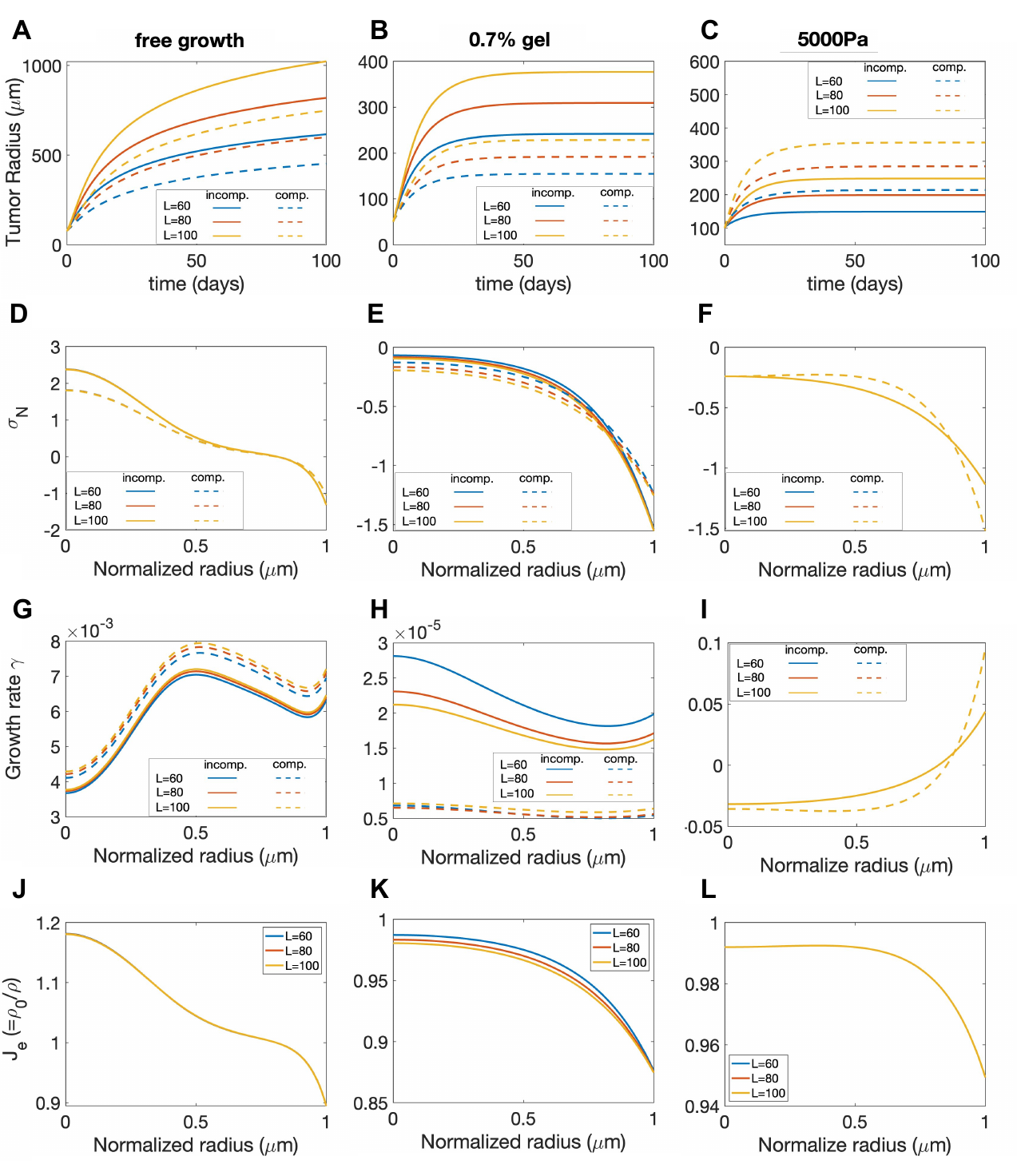}}
    \caption{Study of the parameter $L=\sqrt{D}$,  which characterizes the diffusion length $(D = L^2)$ in our nutrient transportation equation for free case, with gel confinement ($0.7\%$ gel), and with an applied static external pressure ($5000$ Pa). Figures A) $\&$ B) and C) show the evolution of tumor for each case as a function of time (days). Figures D) $\&$ E) and F) indicate the normal stress average $\sigma_N$ distributions in the normalized radial direction at time $T = 100$. Figures G) $\&$ H) and I) indicate the volumetric growth rate distributions in the normalized radial direction at time $T = 100$. Figures J) $\&$ K) and L) show the elastic volumetric variations at time $T = 100$ for the compressible model. The solid lines indicate incompressible results and dashed lines indicate compressible results.}
  \label{ParameterStudyL}
\end{figure}

\end{document}